\documentclass[aps,prb,showpacs,reprint, superscriptaddress]{revtex4-1}

\usepackage{graphicx}
\usepackage{color}
\usepackage{amsmath}
\usepackage{bbm}
\usepackage{amssymb}

\usepackage{dsfont}

\usepackage[utf8]{inputenc}	
\usepackage[T1]{fontenc}

\input epsf

\begin{document}

\title{Correlation induced $d$-\emph{wave} pairing in a quantum dot square lattice}
\author{A. Biborski}
\email{andrzej.biborski@agh.edu.pl}
\affiliation{Academic Centre for Materials and Nanotechnology, AGH University of Science and Technology, Al. Mickiewicza 30, 30-059 Krakow,
Poland}

\author{M. P. Nowak}
\email{mpnowak@agh.edu.pl}
\affiliation{Academic Centre for Materials and Nanotechnology, AGH University of Science and Technology, Al. Mickiewicza 30, 30-059 Krakow,
Poland}

\author{M. Zegrodnik}
\email{michal.zegrodnik@agh.edu.pl}
\affiliation{Academic Centre for Materials and Nanotechnology, AGH University of Science and Technology, Al. Mickiewicza 30, 30-059 Krakow,
Poland}

\begin{abstract}
We consider an electrostatically induced square lattice of quantum dots and study the role of electron-electron correlations in the resulting electronic features of the system. We utilize the Wannier functions methodology in order to construct  Hamiltonian for interacting fermions and find that the change of the depth of the quantum dot confining potential results in a transition from a moderately-, to strong-correlated regime of the system. We obtain the approximate ground state by means of Variational Monte-Carlo method for a wide range of dopings. The values of microscopic parameters, charge gap as well as spin- and pair-correlation functions obtained in the strongly-correlated regime signify the presence of antiferromagnetic spin-ordering and the realization of the Mott insulator phase. Moreover, we report on a two dome structure of the emerging $d$-wave paired state residing on both sides of the half filled case. The obtained results are discussed in view of the well known family of unconventional superconducting materials such as copper based compounds.
\end{abstract}

\maketitle


\section{\label{sec:introduction} Introduction}


Unconventional phases which appear in the correlated electron systems have gathered a significant amount of interest over the years. At low energies, correlated materials show conflicting tendencies towards different symmetry-broken states leading to complex phase diagrams and exotic physical properties such as unconventional forms of superconductivity, magnetism, non-Fermi liquid behavior, and Mott physics\cite{Imada}. In spite of extensive experimental and theoretical effort the complete theoretical description of many prominent examples of correlated system, such as cuprates\cite{Ogata_2008,Agterberg} or heavy-fermion materials\cite{Steglich}, still remains unclear.
One of the main issues in formulating the proper theoretical approach is related to incorporating the electron-electron interaction with satisfactory precision. Calculation methods dedicated to correlated systems are characterized by a high degree of complexity, which limits their applicability only to simplified models and/or systems with a significantly reduced size. In this respect, the seminal Hubbard\cite{Hubbard,Imada,Qin2} and $t$-$J$\cite{Spalek_2007,Imada} models or their derivatives are usually considered as candidates which can allow clarifying the nature of the strongly correlated phenomena.

At the same time, novel correlated systems are being discovered, which can allow for better experimental verification of the proposed theoretical concepts. 
It is believed that the recently synthesized twisted bilayer graphene\cite{Cao1,Cao2} due to its high degree of tunability, will allow answering open questions related to the interacting electron phenomena. The major advantage of twisted bilayer graphene with respect to the previously known correlated systems is that the electron density can be easily tuned by using electronic gates, thus avoiding the disruptive effects of chemical doping. Moreover, the strength of correlations can be controlled by changing the twist angle. Another system in which many microscopic parameters can be controlled experimentally consists of ultracold atoms trapped in an optical lattice\cite{Esslinger,Hofstetter}. In spite of the fact that the latter has been discovered already some time ago, it is still considered as one of the most promising experimental setups to simulate the behavior of the Hubbard model.

Moreover, the  electronic correlations may also play an important role in the physics of nanoscopic systems. Specifically, quantum dots (QDs) are often considered as artificial atoms\cite{Tarucha} and can be regarded as building blocks for more complex devices in which the precise inclusion of electron-electron interactions is indispensable for proper description.
Here, we consider the use of QDs as a route to realize highly tunable strongly correlated electron systems. The advancement in nanofabrication methods allows for the creation of electrostatically controlled QDs\cite{Hendrickx}, as well as their patterns in two-dimensions\cite{Hensgens,vandiepen2021quantum,Mukhopadhyay}. By proper architecture of gates, such a pattern of QDs can serve as an experimental realization of the two-dimensional Hubbard model with both the electron concentration and strength of correlations controlled in situ by means of gate voltage\cite{Byrnes,Byrnes2}. This provides an opportunity to clarify many of the long-standing problems related to emergent phenomena in interacting electron systems.

In this theoretical analysis, we elucidate the phenomena arising from electron-electron interactions by considering a square lattice consisting of electrostatically fabricated dots and use Variational Monte Carlo (VMC) method to capture the correlation effects. We start from an approach complementary with respect to the previously performed studies\cite{Byrnes,Byrnes2}, i.e., we use the real-space picture, and consequently build a translationally invariant system. Notably, we take into account the screening effects resulting from the proximity of the metallic top gate, which is important to reduce the amplitude of two-body interactions \cite{Byrnes} as pointed out by Byrnes et al. Our main finding is that the QDs lattice realizes correlation-driven antiferromagnetic ordering, Mott insulating phase and most importantly $d$-wave superconductivity. So far, the presumption of the existence of $d$-wave superconductivity in such systems has been made only based on the relation between values of interaction amplitudes and the single particle spectrum for the effective interacting models describing QDs lattice\cite{Byrnes}. In this work \emph{we inspect explicitly} if such a scenario takes place and provide theoretical evidence of pairing in the $d$-wave channel for the low-carrier density (implicitly assumed by considering only the lowest band) and within general interacting Hamiltonian. The obtained results are discussed in view of the well known families of unconventional superconducting materials such as cuprates.

In the next section, we describe the model of the QD layer. Subsequently, the section devoted to the methodology covers both the construction of Wannier basis and the calculation of transfer(hopping) and interaction integrals, as well as the solution of many-body Hamiltonian in terms of VMC method. In Sec. III, we present the results obtained for the selected lattice spacing and different depths of the confining potential $V^C$.  We discuss the electronic properties of the system for the selected confinement as a function of charge doping (carrier density) in terms of the analysis of one- and two-body correlation functions, discussing the spin order and the development of charge-gap, as well as, the evidence of singlet pairing. Finally, we summarize the obtained results and conclude them in the very last section.

\section{\label{sec:model}Model}
\begin{figure}
\includegraphics[width=0.40\textwidth]{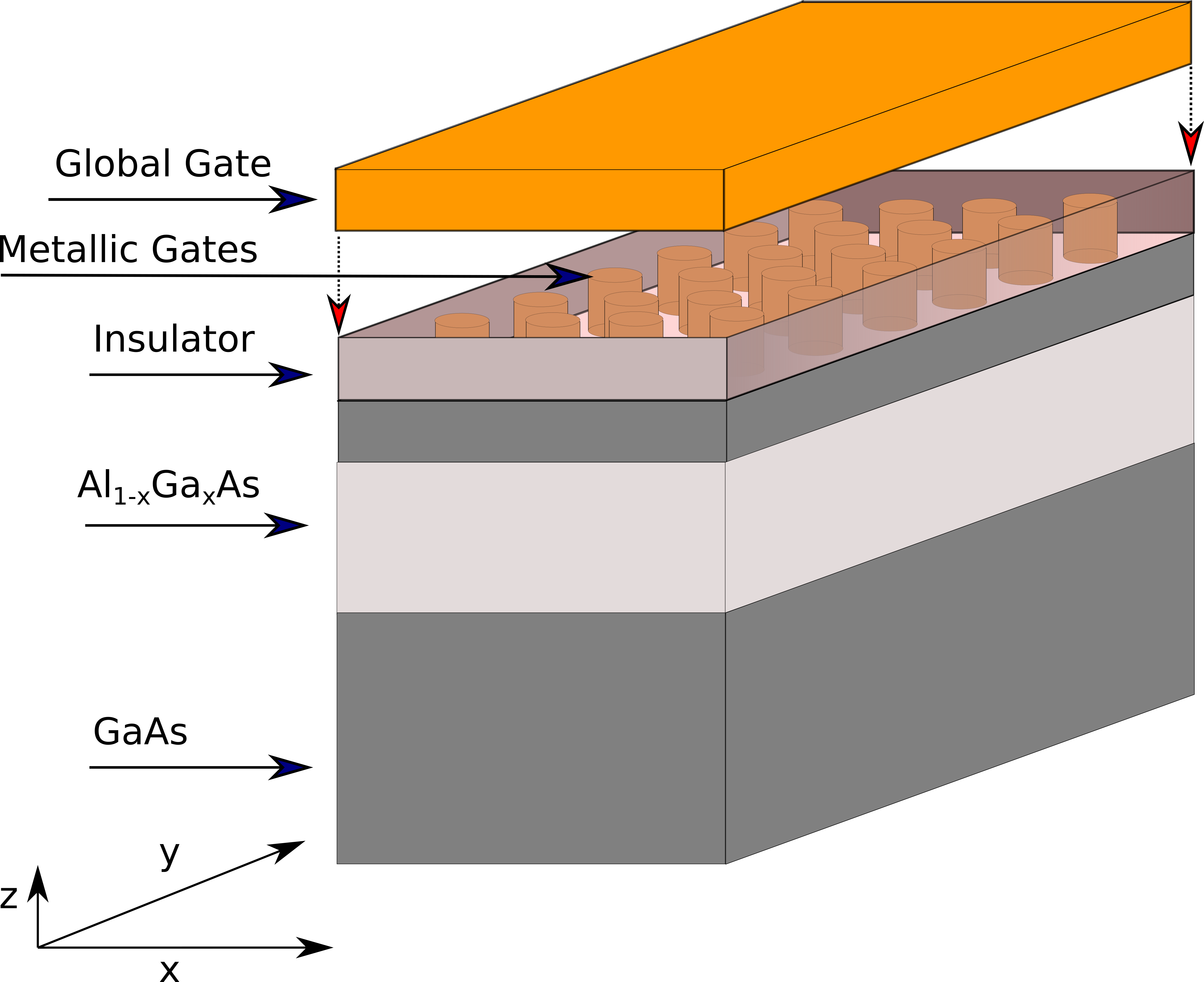}
\caption{\label{fig:device}Schematic representation of the considered structure. Note, that space between the insulator layer and the global gate is only for presentation purposes.}
\end{figure}

We follow the concept of the device presented by Byrnes et al. \cite{Byrnes}, based on $\text{GaAs}\text{/}\text{AlGaAs}$ layered heterostructure (Fig.~\ref{fig:device}) in which two-dimensional electron gas (2DEG) is formed. The top global gate controls the electron concentration in the system, whereas the metallic electrodes, assembled as a square lattice pattern, are immersed in an insulator layer allowing for modulation of the periodic confinement. In this manner, both doping and electron confinement are tunable, allowing us to switch between diverse regimes characterized by different strengths of the electronic correlation as well as scan the resulting phase diagram.

Here, instead of considering the globally assumed periodic potential\cite{Byrnes}, we start from the single-particle spectrum of the isolated QD. A variety of confinement potentials $V^{QD}$ have been exploited to realistically model the electrostatically induced QD\cite{Bednarek2,Ciurla}, e.g., parabolic, Gaussian, power-exponential. We restrict our choice to the form  which exhibits radial symmetry in the $\mathbf{\hat{x}}-\mathbf{\hat{y}}$ plane, as well as, is spatially convergent, i.e., $\text{lim}_{\mathbf{r}\rightarrow\infty}V^{QD}(\mathbf{r}-\mathbf{R})=const$, where $\mathbf{r}=(x,y)$ and $\mathbf{R}$ refers  $V^{QD}$ center. The explicit form of the exploited Gaussian potential for QD is given as
\begin{equation}
    V^{QD}_{\mathbf{R}}(\mathbf{r})=V_0\times\text{exp}\Bigg[-\Bigg(\frac{||\mathbf{r}-\mathbf{R}||}{r_0}\Bigg)^{2}\Bigg],
    \label{eq:vqd}
\end{equation}
where  $V_0$ tunes quantum well depth and $r_0$ controls its planar size. The Gaussian form of $V^{QD}_{\mathbf{R}}$ is believed to describe properly the shape of potential for GaAs based devices for the relatively small electrode radius (less than $10^{2}\text{ nm}$) which in our case is also very close to the value of $r_0$\cite{Bednarek2}. The planar confinement $V^{C}$ originating from the whole pattern of electrodes has the form
\begin{equation}
    V^{C}_{a}(\mathbf{r})=V_{0}-\sum_{i}\sum_{j}V^{QD}_{\mathbf{R_{ij}}}(\mathbf{r}),
\end{equation}
where $a$ is the lattice spacing and $\mathbf{R_{ij}}=(i\times a\mathbf{\hat{x}},j\times a\mathbf{\hat{y}})$. Note, that when $a$ is sufficiently larger than $r_0$, this parametrisation leads to $V^{C}_{a}(i\times a\mathbf{\hat{x}},j\times a\mathbf{\hat{y}})\approx 0$ and $V^{C}_{a}(\frac{i}{2}\times a\mathbf{\hat{x}},\frac{j}{2}\times a\mathbf{\hat{y}})\approx V_0$. 

\section{Method}
We employ the following stages in the computational procedure: (\emph{i}) Schr\"{o}dinger equation solution for assumed  $V^{QD}$; (\emph{ii}) Wannier basis construction; (\emph{iii}) formulation of many body Hamiltonian; (\emph{iv}) VMC solution of the interacting system. We briefly describe each stage in the following subsections.

\subsection{Spectrum of an isolated QD}
According to the form of the confining potential adapted here (Eq.\ref{eq:vqd}), the single QD is parametrized by $\{V_0,r_0\}$.  Analytical solution of Schr\"{o}dinger equation  for the radially symmetrical power-exponential potential is not known \cite{Ciurla}, thus single-particle eigenequation needs to be diagonalized numerically.
As the aforementioned form of the potential exhibits radial symmetry,  the time-independent Schr\"{o}dinger equation can be factorized in a regular manner, i.e., as a product of the radial and angular parts, namely
\begin{equation}
\Theta_{nl}(r,\phi)=R_{nl}(r)\Phi_{l}(\phi).
\end{equation}

Angular part $R_{nl}(r)$ is the solution of the equation
\begin{equation}
\Bigg[r^2\frac{\partial^2}{\partial r^2}+r\frac{\partial}{\partial r} + r^2\frac{2m^{*}}{\hbar^2}\big(E_{nl}-V(r) \big)-\frac{l^2}{\hbar^2}\Bigg]R_{nl}(r)=0,
\label{eq:qdradial}
\end{equation}
and $\Phi_l(\phi)$ is given as
\begin{equation}
\Phi_l(\phi)=\exp(i\frac{l}{\hbar}\phi),
\end{equation}
where $n=0,1,2,...; l=0,\pm 1,\pm 2 ...$ are the principal and angular quantum numbers, respectively. We set  $m^{*}=0.067m_e$, i.e., the value of effective mass for GaAs. Equation (\ref{eq:qdradial}) can be transformed to the form of an eigenproblem for a given value of $l^2$ which can be efficiently solved within the numerical scheme recently proposed by Laliena and Campo \cite{Laliena}, as we implemented in our code.

\subsection{Wannier basis}
The electron wave function in the lattice of QDs can be defined in the basis of Wannier functions constructed with the use of the previously specified $\{\Theta_{nl}(r,\phi)\}$ functions. Although neither $l$ nor $n$ are good quantum numbers for the finite lattice spacing $a$, we apply them for indexing also in the resulting Wannier functions. Namely, the latter are defined as a linear combination of $\Theta_{nl}(r,\phi)$ as
\begin{equation}
    w^{R_{ij}}_{nl}(\mathbf{r})\equiv\sum_{n',l'}\sum_{\mathbf{r_Z}}\alpha_{r_Z,n',l'}\Theta_{n'l'}(\mathbf{r}-\mathbf{R_{ij}}+\mathbf{r_Z}),
    \label{eq:wannier}
\end{equation}
where $\text{lim}_{a\rightarrow\infty}w^{R_{ij}}_{nl}(\mathbf{r})=\Theta_{nl}(\mathbf{r}-\mathbf{R_{ij}})$ condition defines the values of $n,l$ for the elements of the Wannier basis. The coefficients $\{\alpha_{r_Z,n,l}\}$ are to be determined where $\mathbf{r_Z}$ are vectors pointing to $Z$-th nearest neighbouring lattice site. Finite number of orbitals are  taken into account to perform numerical orthogonalization of Wannier functions. Therefore, we consider only those $\Theta_{nl}$ for which $E_{n,l}<V_0$ in the expansion in Eq.~\ref{eq:wannier}. This basis truncation is adequate when $a$ is sufficiently large in comparison to $r_0$, i.e., when the application of the Wannier wave functions is justified. Also, we take into account finite number of neighboring sites, and consider finite, square cluster of size $2m+1 \times 2m +1$.  Next, $\alpha_{r_Z,n',l'}$ are obtained by means of L\"owdin  orthogonalization\cite{Aiken}. When $m$ is systematically increased, central (i.e., those centered at $\mathbf{R}_{mm}=(m\times a,m\times a )$) Wannier functions may be treated as transitionally invariant within the desired numerical precision\cite{Biborski1}.

We assume an infinite quantum well confinement in $\hat{\mathbf{z}}$ direction. Therefore, we formulate the final form of a single particle wave function as a product of $w^{R_{ij}}_{nl}(\mathbf{r})$ and the ground state of the confinement in the direction perpendicular to the lattice, namely
\begin{equation}
    \widetilde{w}^{R_{ij}}_{nl}(\mathbf{r},z)=w^{R_{ij}}_{nl}(\mathbf{r})\sqrt{\frac{2}{L}}\cos\Big(\frac{\pi z}{L}\Big),
\end{equation}
where we take $L=10\text{ nm}$---which is a reasonable value for 2DEG quantum well width\cite{QinJian}. Eventually, we have
\begin{equation}
    \Big\langle \widetilde{w}^{R_{ij}}_{nl}(\mathbf{r},z)\Big|\widetilde{w}^{R_{pq}}_{n'l'}(\mathbf{r},z)\Big\rangle \approx \delta_{R_{ij},R_{pq}}\delta_{nl,n'l'}.
    \label{eq:orthogonality}
\end{equation}
The above approximation becomes exact for $m\rightarrow\infty$. For the localized system, with a relatively small value of  $m$ (i.e., $m < 10$) a nearly (limited by the numerical precision) exact fulfilment of the orthogonality relation occurs, as will be shown further on.

\subsection{Many body Hamiltonian}
In the following, we employ the description of the interacting electron system by means of the second quantization formalism. For the sake of clarity, we map $R_{ij}$ positions to lattice site indices, labeled by $\{i,j,k,l\}$, and, $(l,n)$ to a single band index $\{\mu,\nu,\gamma,\tau\}$. The full electronic Hamiltonian is given by
\begin{equation}
\begin{split}
    \mathcal{\hat{H}}=\sum_{i,\mu,\sigma}\epsilon_{i}^{\mu}\hat{c}^{\dagger}_{i,\mu,\sigma}\hat{c}_{i,\mu,\sigma}+\sum_{i,j}\sum_{\mu,\nu}\sum_{\sigma}t_{ij}^{\mu\nu}\hat{c}^{\dagger}_{i,\mu,\sigma}\hat{c}_{j,\nu,\sigma}+\\ +\frac{1}{2}\sum_{\substack{i,j,\\k,l}}\sum_{\substack{\mu,\nu,\\\gamma,\tau}}\sum_{\sigma,\sigma'}V_{ijkl}^{\mu\nu\gamma\tau}\hat{c}^{\dagger}_{i,\mu,\sigma}\hat{c}^{\dagger}_{j,\nu,\sigma'}\hat{c}_{l,\tau,\sigma'}\hat{c}_{k,\gamma,\sigma},
\end{split}
\label{eq:hamiltonian}
\end{equation}
where $\hat{c}^{\dagger}_{i\mu,\sigma}$ ($\hat{c}_{i,\mu,\sigma}$) are the fermionic creation (annihilation) operators for particles with $\sigma=\{\uparrow,\downarrow\}$. The single particle amplitudes $\{\epsilon_{i^\mu},t_{ij}^{\mu\nu}\}$, as well as two-body $V_{ijkl}^{\mu\nu\gamma\tau}$interactions are to be determined by means of calculating the following matrix elements
\begin{equation}
    t_{ij}^{\mu\nu}=\Bigg\langle w_{\mu}^{i}(\mathbf{r})\Bigg|-\frac{\hbar^2}{2m^{*}}\nabla^{2}_r+V_a^{C}(\mathbf{r})\Bigg|w_{\nu}^{j}(\mathbf{r})\Bigg\rangle,
    \label{eq:hoppingamp}
\end{equation}
where we intentionally disregard the integration with respect to $z$, since for a given $\{V_0,r_0,L,d\}$ this only shifts the diagonal elements (i.e., $t_{ii}^{\mu\mu}=\epsilon_{i}^{\mu}$) by a constant value $\hbar^2\pi^2/2m^{*}L^2$. 

The two-body interaction terms take the form\cite{Spalek_EDABI,Biborski1,DRN,Byrnes}
\begin{equation}
\begin{split}
    V_{ijkl}^{\mu\nu\gamma\tau}= \Big\langle \widetilde{w}_{\mu}^{i}(\mathbf{r},z)\widetilde{w}_{\nu}^{j}(\mathbf{r'},z')\Big|\hat{V}_{e-e}\Big|\widetilde{w}_{\gamma}^{k}(\mathbf{r},z)\widetilde{w}_{\tau}^{l}(\mathbf{r'},z')\Big\rangle.
    \end{split}
    \label{eq:interamp}
\end{equation}
Note that for the considered system the electrostatic screening resulting from the metallic gates should be included\cite{Byrnes}, what leads to the following form of the electron-electron interaction, $V_{e-e}$
\begin{equation}
    \hat{V}_{e-e}(\mathbf{r},\mathbf{r'},z,z')=\frac{e^2f(\mathbf{r},\mathbf{r'},z,z')}{4\pi\varepsilon_0\varepsilon\sqrt{|\mathbf{r}-\mathbf{r'}|^2+(z-z')^2}},
    \label{eq:iintegrals}
\end{equation}
where 
\begin{equation}
f(\mathbf{r},\mathbf{r'},z,z')=1-\frac{\sqrt{|\mathbf{r}-\mathbf{r'}|^2+(z-z')^2}}{\sqrt{|\mathbf{r}-\mathbf{r'}|^2+(z+z'+2d)^2}}
\end{equation}
is the screening function and $\varepsilon=12.9$ is the dielectric constant. 

It should be emphasized that $\mathcal{H}$ in Eq. (\ref{eq:hamiltonian}) is given in a general form, since  the quantitative relations among the magnitudes of microscopic parameters are not known a priori. However, the number of terms in the Hamiltonian may be reduced by a posteriori analysis, i.e., after obtaining integrals for the considered system parametrisation, one may exclude the subset of terms in the Hamiltonian, based on the observation of a marginal value (i.e., close to the numerical precision) of the corresponding microscopic amplitudes.

\subsection{Variational Monte-Carlo}

We employ the VMC method\cite{Becca,TOULOUSE} to find the approximate ground state of the interacting system. The absence of the infamous sign problem as well as the relatively high numerical efficiency together with the opportunity to exploit flexible variational ansatzes make this approach very useful for a wide class of fermionic systems for which the consistent inclusion of electronic correlation\cite{Becca} is indispensable. Particularly, fermionic lattice models formulated in the second quantization language can be treated efficiently\cite{Hui-Hai,Kato,Yokoyama,Zegrodnik1,Biborski1,Biborski2} in the framework of VMC. 

The choice of many-body variational ansatz $|\Psi_T\rangle$ is crucial for the valid determination of the ground state properties in VMC. Typically, $|\Psi_T\rangle$ is given as the state resulting from acting of the correlation factor  $\mathcal{\hat P}$ and the projection operator $\mathcal{\hat{L}}$ on the so-called non-interacting wave function $|\Psi_0\rangle$
\begin{equation}
    |\Psi_T\rangle =  \mathcal{\hat P}\mathcal{\hat{L}} |\Psi_0\rangle.
\end{equation}

One of the possible forms of $|\Psi_0\rangle$ is the pair-product or Pfaffian wave function\cite{Misawa,Becca} given as
\begin{equation}
    |\Psi_0\rangle=\Big[\sum_{i,j}\sum_{\mu\nu}\sum_{\sigma,\sigma'}F_{i\mu,j\nu}^{\sigma\sigma'}\hat{c}_{i,\mu,\sigma}^{\dagger}\hat{c}_{j,\nu,\sigma'}^{\dagger}\Big]^{N_e/2}|0\rangle,
    \label{eq:psi0}
\end{equation}
where $F_{i\mu,j\nu}^{\sigma\sigma'}$ are variational parameters to be determined, $N_e$ is a number of electrons in the system, and, $|0\rangle$ refers to the vacuum state. Note that when the summation is narrowed to the case where $\sigma\neq\sigma'$, the so-called anti-parallel wave function is realized which is particularly useful for systems in which the total spin $z$-component is zero \cite{Misawa}. 

Before we briefly describe $\mathcal{\hat{P}}$ and $\mathcal{\hat{L}}$ operators applied in this work, it is suitable to sketch the idea of the sampling scheme. The whole method is founded on the variational principle, i.e.,
\begin{equation}
    E_{G}\leq E_T\equiv\frac{\langle\Psi_T|\mathcal{\hat{H}}| \Psi_T\rangle}{\langle\Psi_T|\Psi_T\rangle},
\end{equation}
where $E_G$ is the ground-state energy. As VMC operates in real space, $|\Psi_T\rangle$ can be expressed in terms of the expansion in the $\{|x\rangle\}$ basis defined explicitly below
\begin{equation}
|x\rangle\equiv|x_{\uparrow}\rangle\otimes|x_{\downarrow}\rangle=\prod_{i,\mu \in x_{\uparrow}}\hat{c}^{\dagger}_{i,\mu,\uparrow}\prod_{j,\nu\in x_{\downarrow}}\hat{c}^{\dagger}_{i,\nu,\downarrow} |0\rangle,
\end{equation}
where $x_{\uparrow}$ and $x_{\downarrow}$ refer to the set of occupied spin-up and spin-down states, respectively. Configurations $|x\rangle$  are sampled according to the probability density function $\rho(x)\propto |\langle x|\Psi_T\rangle|^2$ and trial ground state energy $E_T$ is estimated as 
\begin{equation}
E_T\approx\frac{1}{M}\sum_m^{M}\rho(x_m)\frac{\langle \Psi_T|\mathcal{H}|x_m\rangle}{\langle \Psi_T | x_m\rangle}=\frac{1}{M}\sum_m^{M}\rho(x_m)E_T^{loc}(x_m),
\end{equation}
i.e., as an average of local energy $E_T^{loc}(x)$ over assumed number of samples $M$. The proper selection of $|\Psi_0\rangle$ is one of the most important steps in the variational ansatz construction. However, disregarding the possibility of inclusion of the so-called back-flow correlations\cite{Becca} in $|\Psi_0\rangle$, we capture the correlation effects in a standard manner\cite{Becca, Misawa}, i.e., by acting with a Hermitian operator $\mathcal{\hat{P}}$ 
\begin{equation}
\begin{split}
\mathcal{\mathcal{\hat{P}}}(\{g_{i\mu}\},\{v_{i\mu, j\nu}\},\{\alpha_{n\mu,t\nu}^{d4},\alpha_{n\mu,t\nu}^{h4}\})=\\=\mathcal{\hat{P}}_G(\{g_{i\mu\}})\mathcal{\hat{P}}_J(\{v_{i\mu,j\nu}\})\mathcal{\hat{P}}_{d-h}(\{\alpha_{n\mu,t\nu}^{d4},\alpha_{n\mu,t\nu}^{h4}\})
\end{split}
\end{equation}
on $|x_m\rangle$.

In the above $\mathcal{\hat{P}}_G$ , $\mathcal{\hat{P}}_J$ and $\mathcal{\hat{P}}_{d-h}$ are \emph{Gutzwiller}, \emph{Jastrow} and \emph{doublon-holon} correlators respectively\cite{Becca,Misawa}, and, $\{g_{i,\mu}\}$, $\{v_{i\mu,j\nu}\}$, $\{\alpha_{n\mu,t\nu}^{d4},\alpha_{n\mu,t\nu}^{h4}\}$ are related variational parameters.
While $|\Psi_0\rangle$ defined in Eq. (\ref{eq:psi0}) explicitly describes a constant number of particles, we also apply the projection on $S^z_{tot}=0$ state leading to an anti-parallel form of $|\Psi_0\rangle$.

Subsequently, the trial energy $E_T$ can be minimized within Stochastic Reconfiguration Method, which allows for the efficient optimization with respect to the set of variational parameters\cite{Zegrodnik1,Biborski1,Biborski2}. We perform VMC simulations using a recent, highly efficient, general-purpose package \emph{mVMC}\cite{Misawa} elaborated by Misawa et al.

We also remark that other computational methods suitable for the description of strongly correlated systems may be applied for the solution of the developed interacting Hamiltonian, therefore in the Appendix~\ref{Appendix} we provide the values of the computed integrals.

\section{Results}
In this section, we first describe the details of the single-particle picture on which the interacting model is founded. Subsequently, we focus on the electronic properties of the system obtained by means of VMC calculations.

\subsection{Single particle picture}
\begin{figure}
    \includegraphics[width=0.45\textwidth]{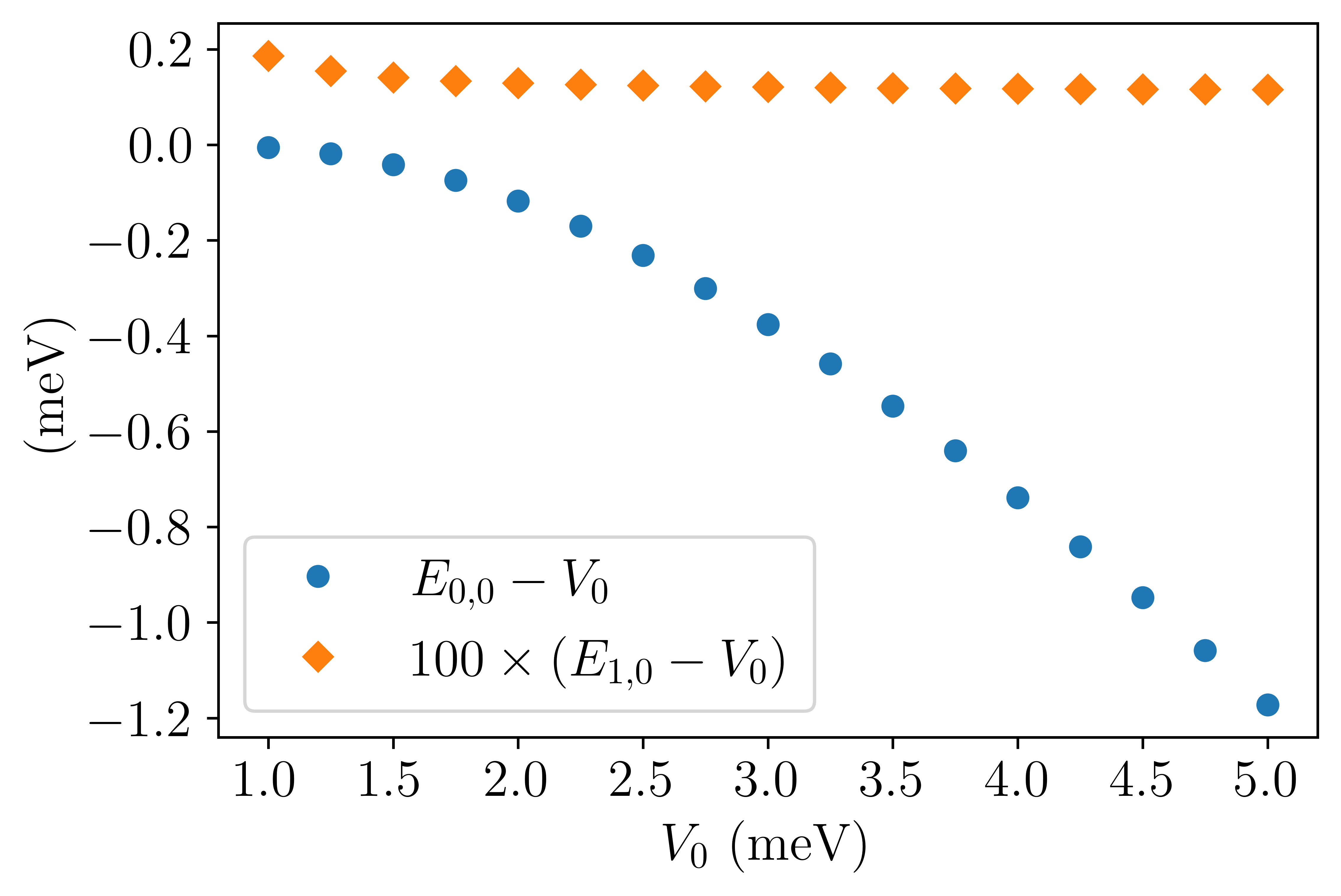}
    \caption{The ground- and the first excited states of isolated QD as a function of $V_{0}$. Note, that only one bound state exists for the considered range of $V_0$. The values corresponding to $E_{01}-V_{0}$ are multiplied by factor of 100 for clarity.}
    \label{fig:figure3}
\end{figure}

The choice of $r_0=20 \text{ nm}$ and $V_0 \in [1.0,5.0]\;\text{meV}$ provides us with the spectrum of isolated QD containing only a single bound state $\Theta_{0,0}$. The energy levels corresponding to the ground, as well as to the first excited state as a function of $V_0$ are presented in Fig. \ref{fig:figure3}. 

\begin{figure}
    \includegraphics[width=0.45\textwidth]{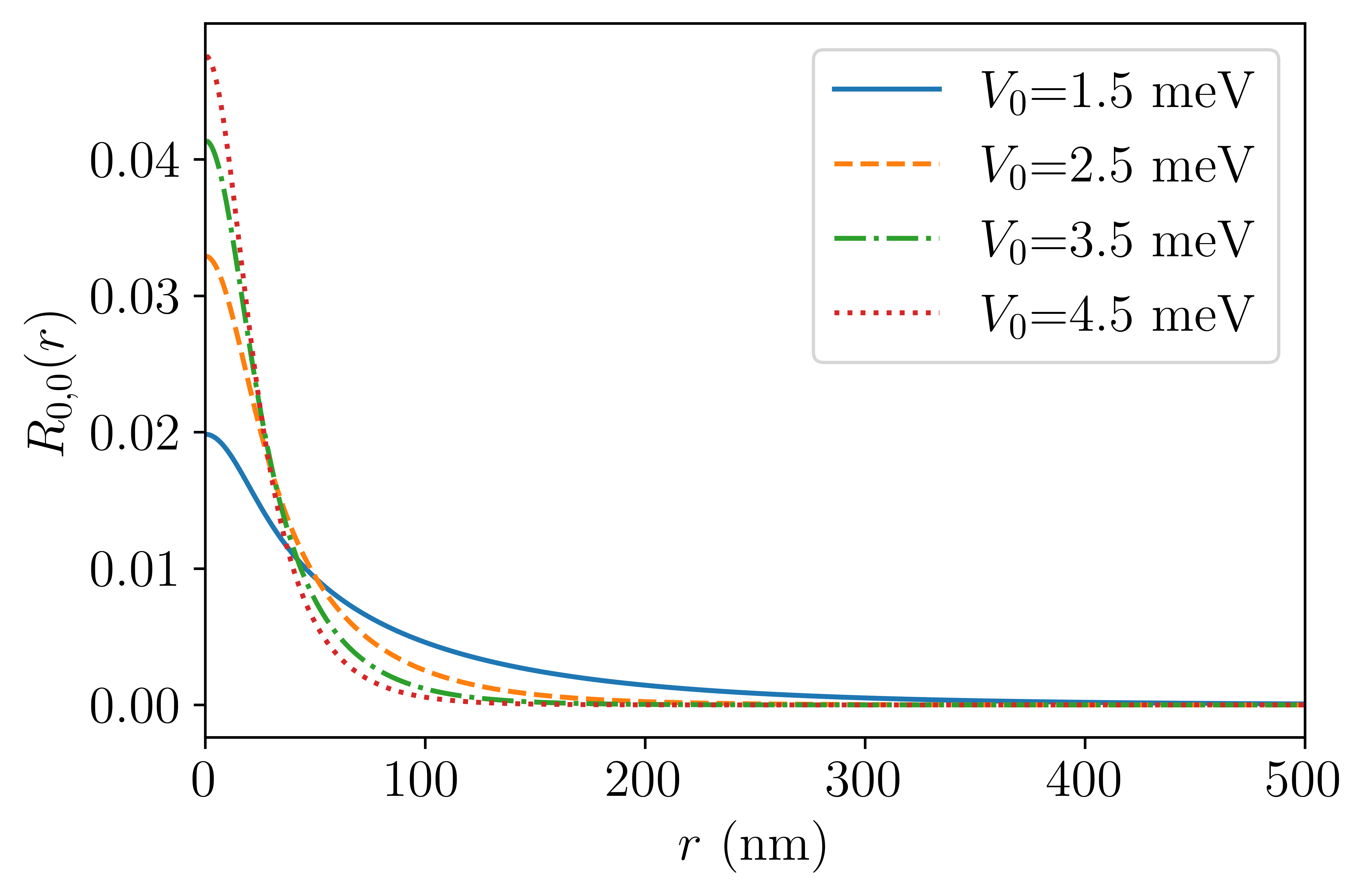}
    \caption{The evolution of radial part $R_{00}(r)$ for $r_0=20\text{ nm}$ and selected values of $V_0$.}
    \label{fig:figure4}
\end{figure}

As expected, by changing $V_0$ we can tune the bound state energy and as a result modify the band structure of the QD pattern as well as the strength of electronic interactions as we show explicitly in the following. For the chosen form of $V_{\mathbf{R}}^{QD}$ potential, $a$ should be sufficiently larger than $r_0$ so that the potentials of subsequent QDs are relatively well separated and do not overlap. On the other hand, $a$ has to be adequately small to allow for non-zero electron hopping and the development of an electronic band characterized by the reasonable width. We identify that $a$ being few (4-5) times larger than $r_0$ fulfills both conditions as legitimated in view of the periodic potential formation, i.e., $V_a^{C}(m\times a/2,n\times a/2)\approx V_0$ holds. In our calculations, we take $a=100\text{ nm}$ and $r_0=20\text{ nm}$. For such a choice the decay of the radial part of the wave function $R_{0,0}$ is presented in Fig.~\ref{fig:figure4}. 

The considered quantum dot lattice, where only the lowest band is occupied and thus the number of electrons per QD is limited, is suitable for quantum transport measurements as the high-mobility 2DEG can be realized already for electron densities substantially below $10^{11}/\text{cm}^{2}$ at $T=0.3\text{ K}$\cite{Chung}. 

As one can see from Fig.~\ref{fig:figure4} for larger values of $V_0$ the decay of  $\Theta_{0,0}(\mathbf{r})$ is faster with increasing $r$. It is unknown a priori what is the reasonable cut-off for $\mathbf{r_Z}$ in Eq.~\ref{eq:wannier}. We find that $Z\lessapprox5$ should be considered as suitable choice for systems with $V_{0}\gtrapprox 2\text{ meV}$, i.e., for those for which the stronger correlation regime is expected. On the other hand for the weaker confinement the $\mathbf{r_Z}$ cut-off has to be significantly expanded. This observation is natural, since for $V_{0}\rightarrow0$ the system reduces to a 2DEG with no QD confinement potentials in $\mathbf{\hat{x}}-\mathbf{\hat{y}}$ plane, thus a highly delocalized basis has to be utilized. 

\begin{figure}
    \includegraphics[width=0.45\textwidth]{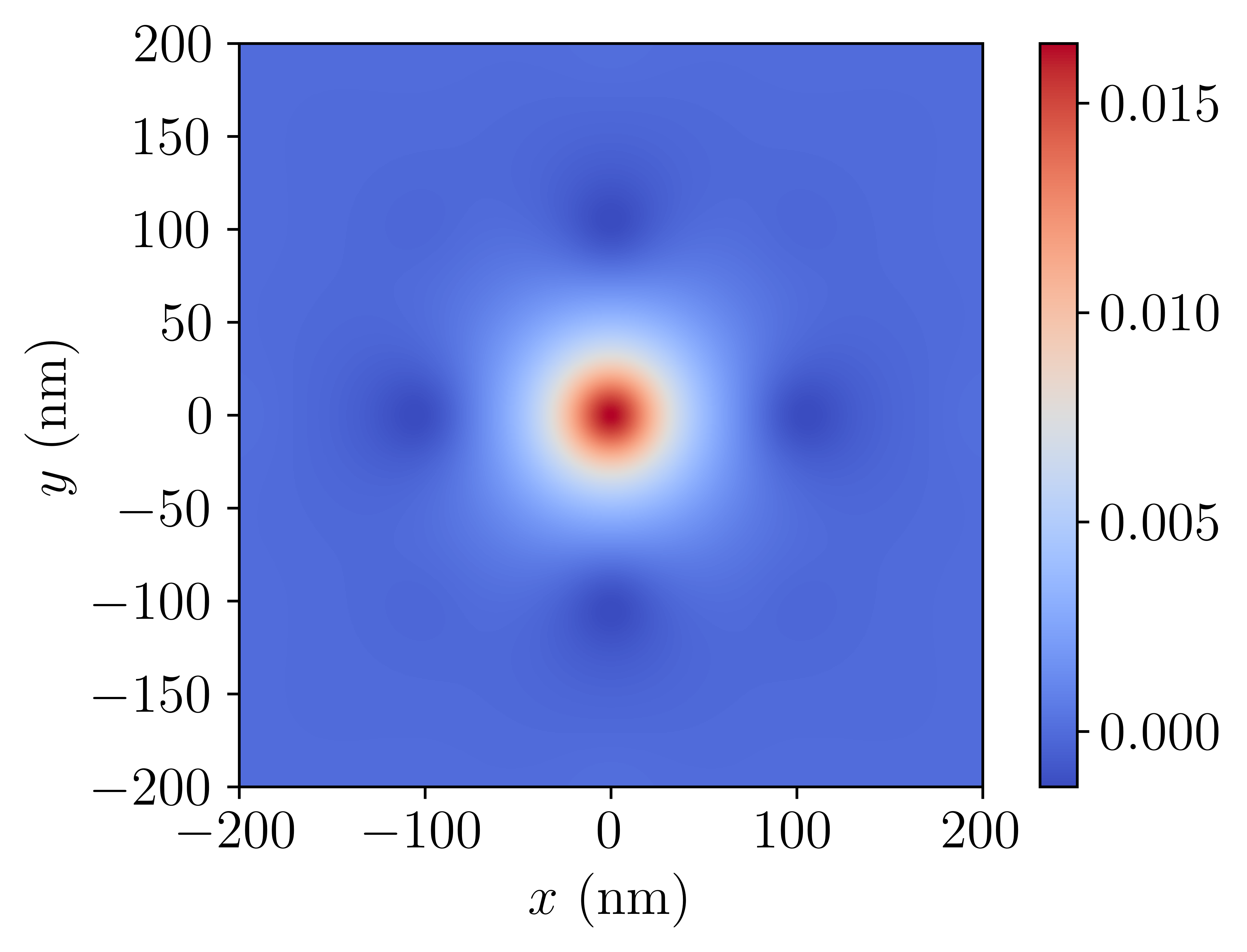}
    \caption{The projection of $w_{00}^{R_{0,0}}(x,y)$---i.e., the radial part of Wannier function---onto the $\mathbf{\hat{x}}-\mathbf{\hat{y}}$ plane for $V_0=2.5\text{ meV}$. }
    \label{fig:figure5}
\end{figure}

\begin{figure}
    \includegraphics[width=0.5\textwidth]{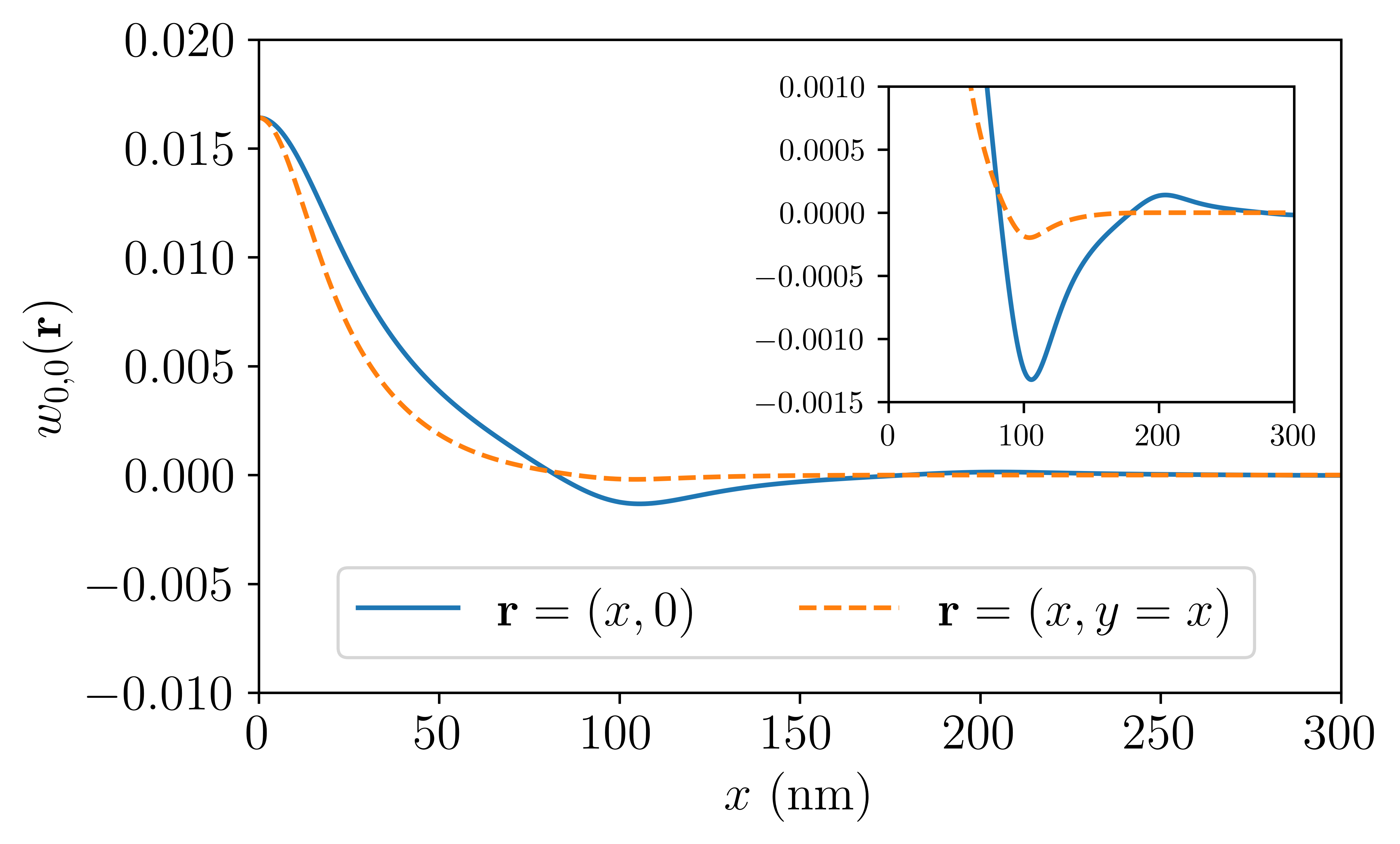}
    \caption{Wannier states $w_{0,0}^{R_{00}}$ obtained for $V_0=2.5 \text{ meV}$ and $a=100\text{ nm}$ as a function of $\mathbf{r}=(x,0)$ and $\mathbf{r}=(x,y=x)$. Well defined local extrema are visible in the vicinity of location of the nearest-, second- and third-nearest sites as shown in the inset.}
    \label{fig:figure6}
\end{figure}

In Figs.~\ref{fig:figure5},\ref{fig:figure6} we show $w_{0,0}^{R_{00}}$ for $V_0=2.5\text{ meV}$ as an example. One finds that the function is well localized in the vicinity of central lattice site. However, local extrema  are also present (see  Fig.~\ref{fig:figure6}) which correspond to $\mathbf r=(a,0)$, $\mathbf{r}=(a,a)$ and $\mathbf{r}=(2a,0)$, i.e, to the first-, second-, and the third-nearest neighbor QDs respectively. The more distant extrema are below the numerical resolution. Subsequently, we validate the whole procedure by inspecting the orthogonality condition given in Eq.(\ref{eq:orthogonality}), i.e., by computing $\big\langle\widetilde{w}_{00}^{R_{00}}\big|\widetilde{w}_{00}^{R_{00}+r_{ij}}\big\rangle$, where $\mathbf{r}_{ij}$ are vectors pointing to the neighboring sites up to the fifth-nearest neighbour.  All the numerical integrations presented in this paper were carried out with the use of the Cuba library\cite{Cuba}. The orthogonality discrepancy does not exceed the value of $\propto10^{-4}$, which we also estimate as the numerical accuracy of Wannier function overlap integrals. Therefore, we find the construction of the Wannier basis as well founded.

\begin{figure}
 \includegraphics[width=0.5\textwidth]{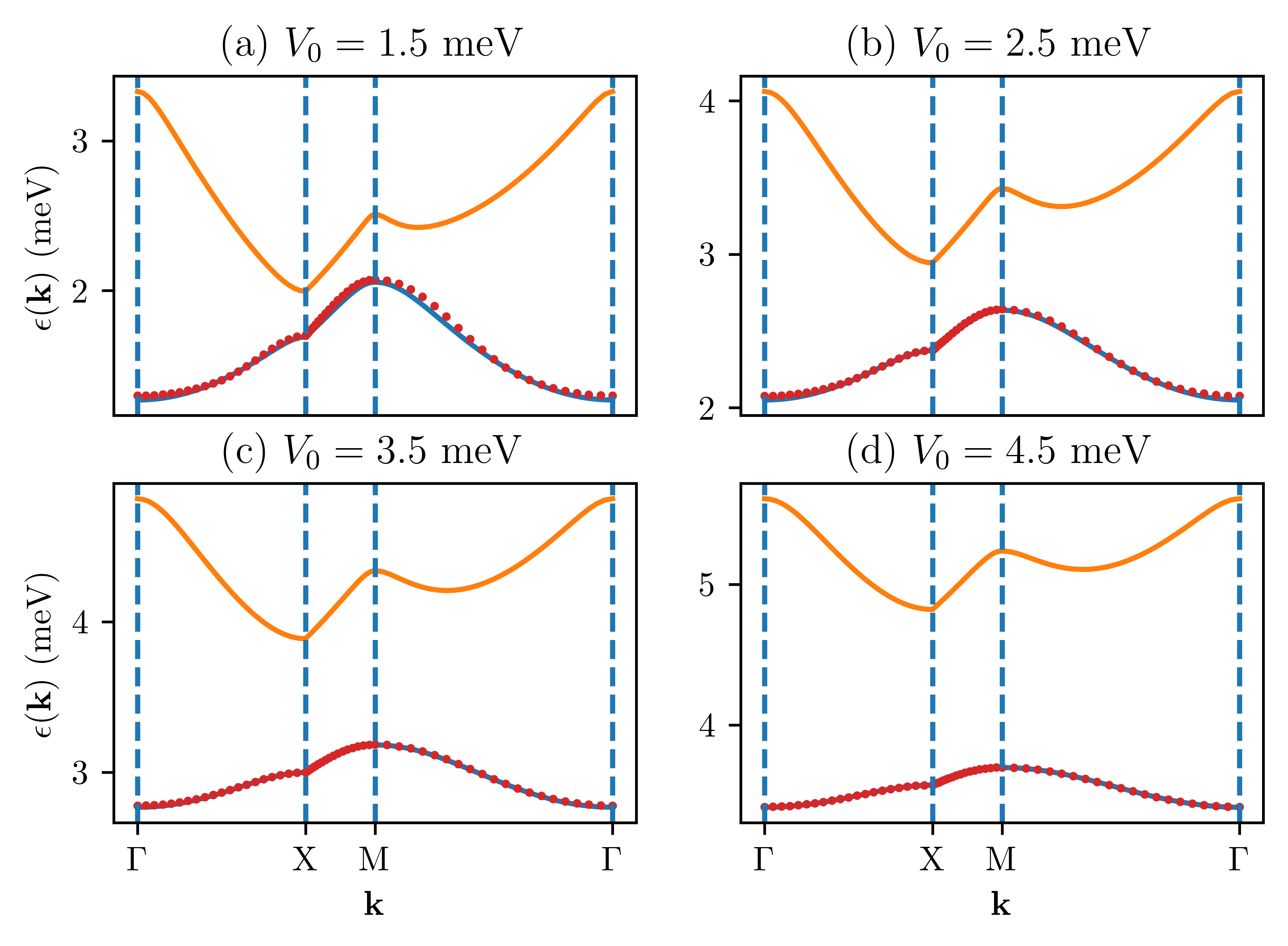}
 \caption{Bare dispersion relations $\epsilon(\mathbf{k})$ resulting from the first two lowest lying bands for the selected values of the confinement potential. The lower bands refer to the constructed Wannier basis (symbols) matching the results obtained numerically for the translationally invariant QD lattice (solid blue lines). Also the band gap enhancement between the first two lowest bands with increasing value of $V_0$ is noticeable.}
 \label{fig:figure7}
\end{figure}

Next, we compare the dispersion relations $\epsilon(\mathbf{k})$ for the non-interacting system, resulting from our Wannier basis, with those computed directly from numerical diagonalization of Schr\"{o}dinger equation for a QD layer
within the \emph{Kwant} package\cite{Kwant}. Obviously, such calculation is also biased, e.g., by the finite mesh-density. We consider a discrete computational mesh $100 \text{ nm} \times 100 \text{ nm}$ with discretization constant $\Delta a=1\text{ nm}$ containing a single QD. The mesh is than periodically repeated, effectively describing a translationally invariant, two-dimensional QD lattice. We find very good agreement between both approaches as shown in Fig.~\ref{fig:figure7} where the bare dispersion relations $\epsilon(\mathbf{k})$ are plotted for the representative set of $V_0$. As supposed, small albeit still noticeable differences are present for $\mathbf{k}=(0,0)$, i.e., for the $\Gamma$ point in the momentum space. As one can see for the stronger confinement an indirect gap opens in the band structure. For $V_0=2.5\text{ meV}$ the gap is about $0.5 \text{ meV}$ and increases  with increasing $V_0$.

\begin{figure}
 \includegraphics[width=0.5\textwidth]{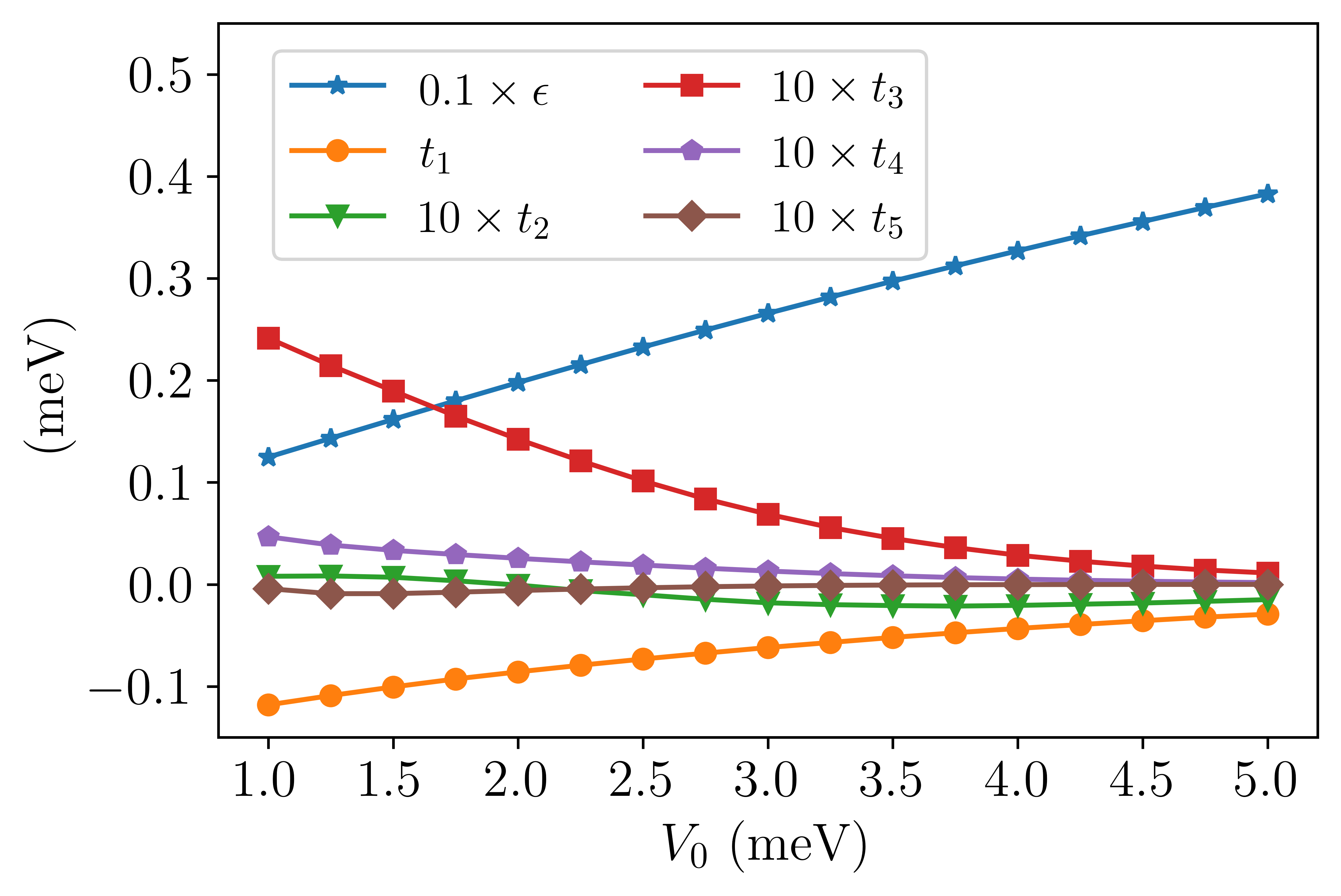}
 \caption{The evolution of hopping integrals as a function of $V_0$. Note, that value referring to $\epsilon$ is divided by factor of $10$, whereas values of $t_{Z}$ related to $Z>1$ are  multiplied by the same factor for the sake of clarity.}
 \label{fig:figure8}
\end{figure}

For the sake of completeness, in Fig.~\ref{fig:figure8} we present the values of hopping integrals for the considered range of $V_0$. We relabel each calculated integral $t_{i(Z),j(Z)}^{00}\equiv t_Z$, in such a way that the pair $(i,j)$ of lattice sites correspond to the $Z$-th nearest neighbor. Namely, for $Z\leq=5$ the sets of corresponding  vectors $\mathbf{R}_{ij}$ which connect $i$-site with $j-site$ are given as
\begin{subequations}
\begin{align}
   Z=1:&\text{ }\mathbf{R}_{ij}\in\{(\pm a,0),(0,\pm a), \}\\
   Z=2:&\mathbf{R}_{ij}\in\{(\pm a,\pm a),(\mp a,\pm a) \},\\
   Z=3:&\mathbf{R}_{ij}\in\{(\pm 2a,0),(0,\pm 2a) \},\\
   Z=4:&\mathbf{R}_{ij}\in\{(\pm 2a,\pm a),(\mp 2a,\pm a),(\pm a,\pm 2a),(\mp a,\pm 2a) \},\\
   Z=5:&\mathbf{R}_{ij}\in\{(\pm 2a,\pm 2a),(\mp 2a,\mp 2a) \}.
\end{align}
\label{eq:Z_number}
\end{subequations}
The absolute value of the hopping integral to the nearest-neighbor is the dominant one, as expected. The characteristic hopping energies correspond to temperatures $|t_1|/k_B\sim 0.1-1\text{K}$ (for $V_0\sim 1 - 3.5\text{ meV}$), which are reachable by the modern dilution refrigerators. The hoppings referring to the more distant neighbors are substantially smaller (c.a. $\times10$). We find $|t_3| > |t_2|$ for the whole considered range, and, $|t_4| > |t_2|$ for $V_0\gtrapprox 3 \text{ meV}$. This can be caused by the particular choice of the gauge implicitly encoded in the adopted procedure for the generation of  Wannier basis. Notably, the obtained Wannier basis reproduces the band structure obtained directly in the momentum space (cf. Fig.\ref{fig:figure7}). 

Moreover, as one may deduce from Fig.~\ref{fig:figure8}, for $Z>0$ all amplitudes of hopping integrals taken into account decay with increasing $V_0$. It is a natural consequence of the stronger localization of Wannier functions with increasing planar confinement. Finally, the role of $t_5$ is marginal---its absolute value does not exceed $6\times10^{-4}\text { meV}$.

\subsection{Electronic interactions and many-particle Hamiltonian}
Whereas the single particle energy spectrum for a given value of $L$ is only shifted by the ground state energy of infinite confinement in $\mathbf{\hat{z}}$ direction, it is not the case for the interaction amplitudes defined in Eq. (\ref{eq:iintegrals}). Furthermore, the thickness of the spacer layer $d$ impacts  the magnitude of electronic interactions and can be regarded as a parameter to be tuned in the experimental setup\cite{Byrnes2}. 

\begin{figure}
 \includegraphics[width=0.5\textwidth]{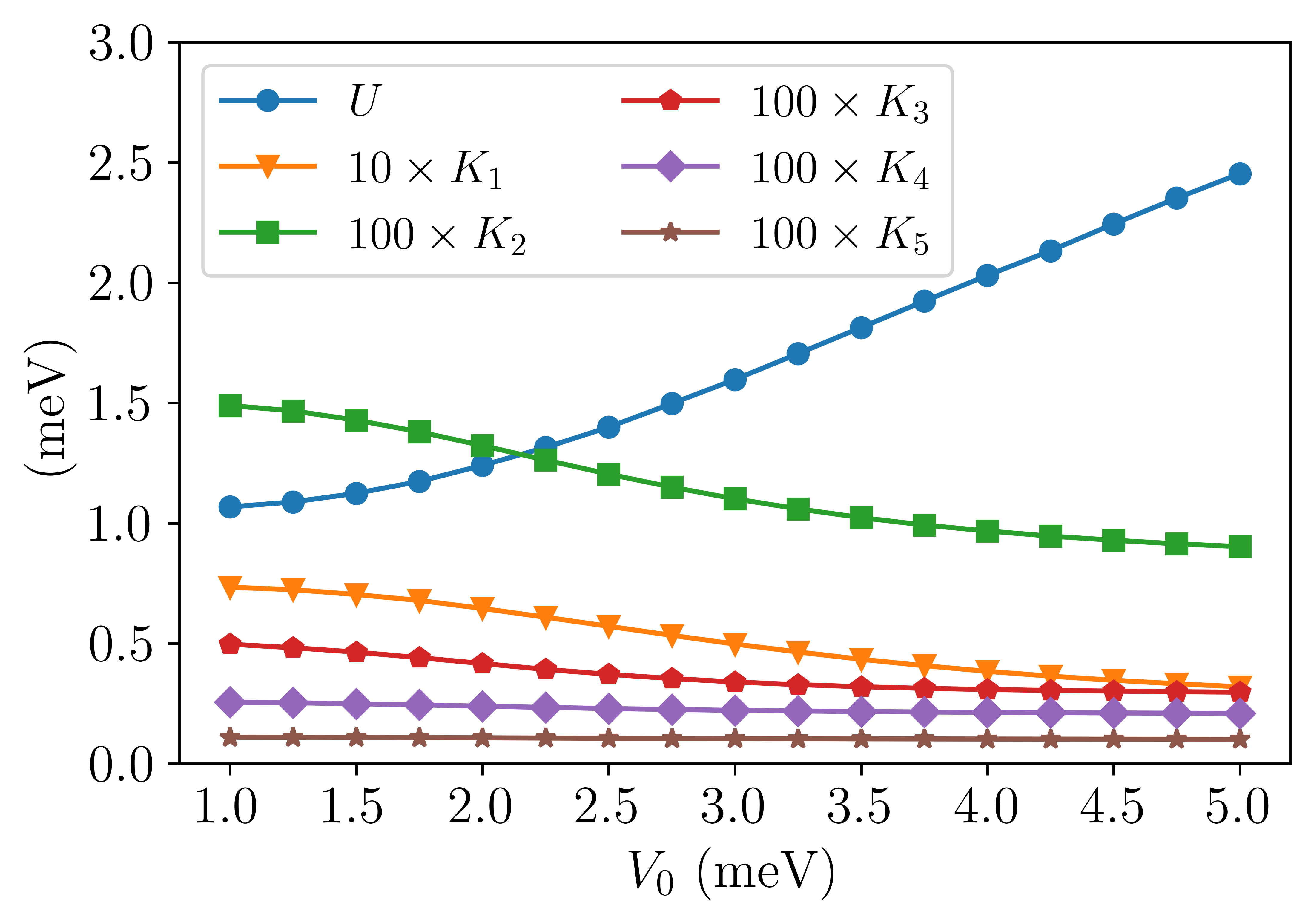}
 \caption{Density-density interactions as a function of $V_0$ for $d=10 \text{ nm}$. The dominant role of intra-site Hubbard interaction $U$ is clearly evidenced.}
 \label{fig:figure9}
\end{figure}

\begin{figure}
 \includegraphics[width=0.5\textwidth]{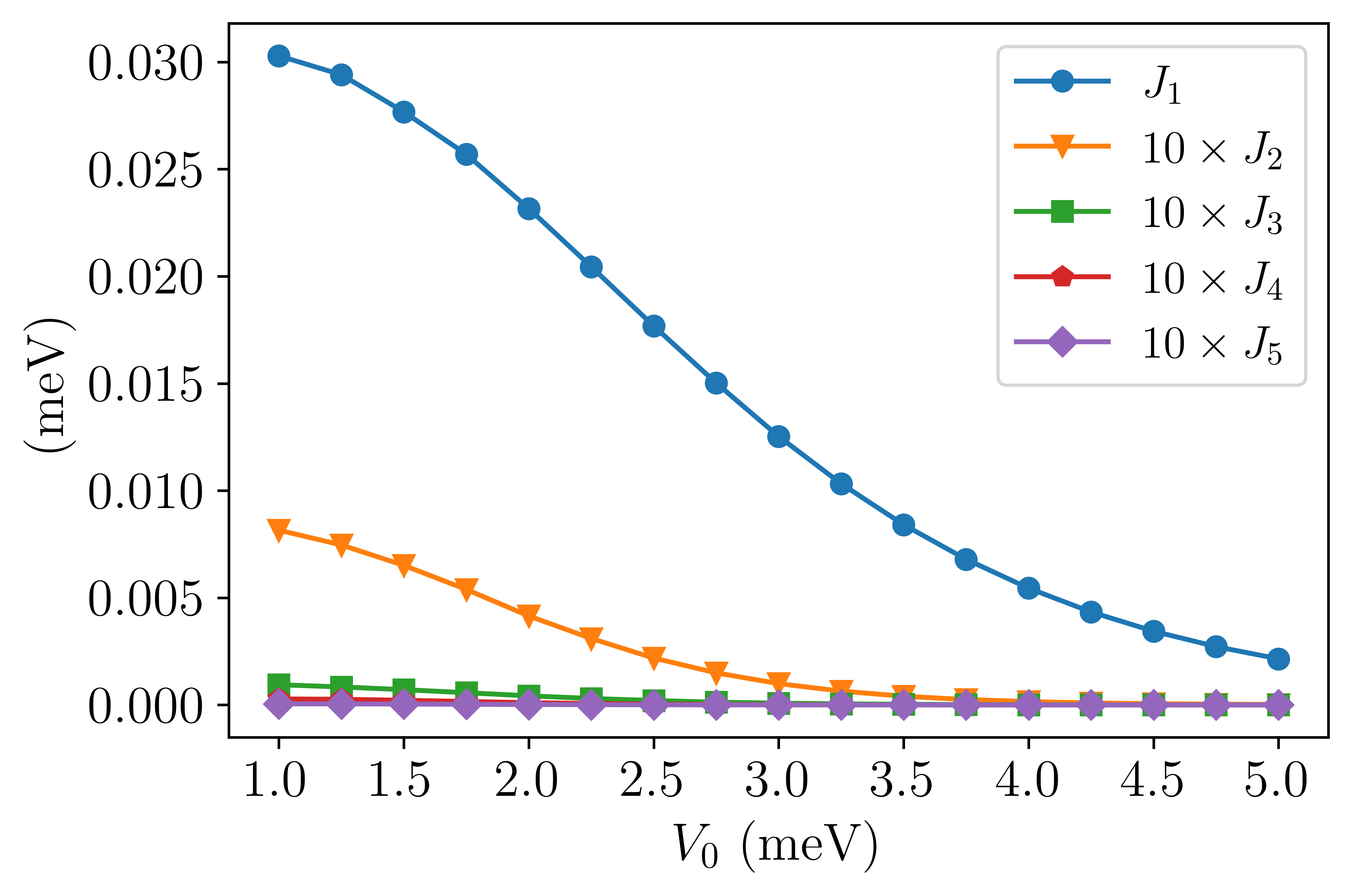}
 \caption{Exchange integral amplitudes $J$ as a function of $V_0$ obtained for the screening parameter $d=10\text{ nm}$. All values are positive therefore refer to tendency of ferromagnetic order. }
 \label{fig:figure10}
\end{figure}

\begin{figure}
 \includegraphics[width=0.5\textwidth]{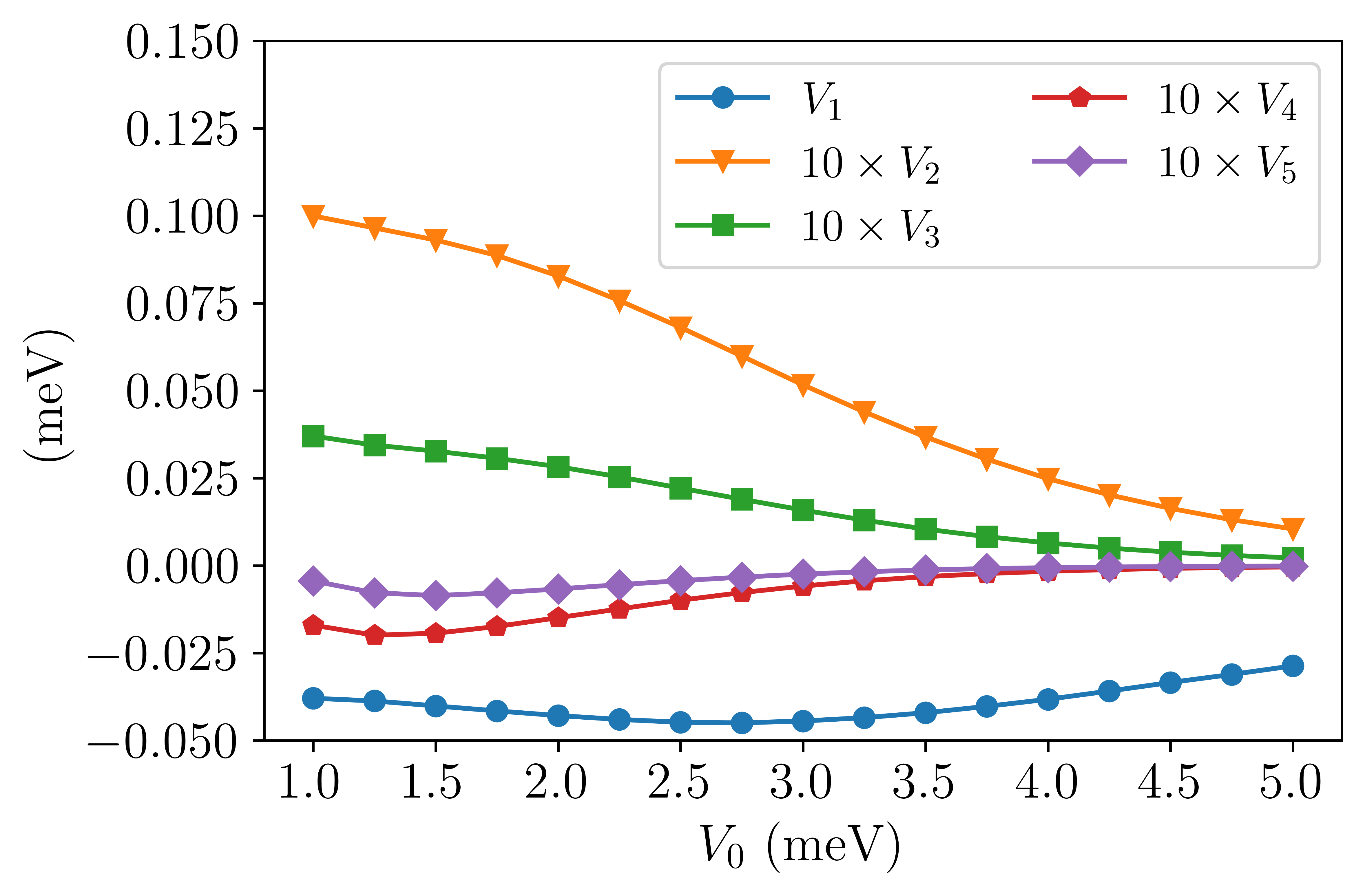}
 \caption{Correlated-hopping amplitudes as a function of $V_0$ for $d=10 \text{ nm}$. Note, that $|V_1|$ is of order of $K_1$ and its dependence on the gate potential value is non-monotonical in the considered range of $V_0$.}
 \label{fig:figure11}
\end{figure}

In Figs.~\ref{fig:figure9}-\ref{fig:figure11} we present the values of two-center interactions obtained for $d=10\text{ nm}$. We use the following notations for each type of integrals, omitting band labels, since we consider only a single lowest lying band as presented above. Namely, we define
\begin{subequations}
\begin{align}
   V_{iiii}&\equiv U,\\
   V_{ijij}&\equiv K_{ij},\\
   V_{iijj}&=V_{ijji}\equiv J_{ij},\\
   V_{ijjj}&=V_{jijj}=V_{jjij}=V_{jjji}\equiv V_{ij}.
\end{align}
\end{subequations}
One identifies $U$ as the onsite Coulomb repulsion (the so-called Hubbard term), whereas  $K_{ij}$ are density-density intersite Coulomb interactions. The exchange-type interactions $J_{ij}$, as well as, the so-called correlated-hopping terms $V_{ij}$, are invariant under the index exchange due to the fact that the obtained Wannier functions are real. We also follow the nomenclature defined for one-body integrals (see Eq.~\ref{eq:Z_number}), e.g., $K_1$ refers to the density-density interaction between the nearest-neighboring QDs.

As we find that the values of three- and four-center interaction integrals are  $\lessapprox 10^{-3}\text{ meV}$, i.e., close to the estimated numerical precision, we disregard their analysis and exclude them from the final form of the interacting Hamiltonian. However, their role can be regarded as quantitatively influential for multiband models as it has been evidenced for isolated QD-like systems\cite{DRN}.

In Fig.~\ref{fig:figure9} we present the values of $K_Z$ up to $Z=5$, where we follow convention assumed also for hopping integrals, e.g., $K_{i(Z),j(Z)}=K_Z$. As expected, the value of $U$ increases with the increasing depth of the confinement potential, becoming dominant for deep potentials. The density-density interaction amplitudes gradually decay as the height of the barrier between QDs is increased. Note that $K_1$ is of the order of $|t_1|$, therefore inter-site interactions are non-negligible and the system has to be described with the use of the Hamiltonian which is more complex than that related to pure Hubbard model. The exchange integrals are positive (see Fig.~\ref{fig:figure10}), thus indicating the tendency for a ferromagnetic ordering. $J_1$ is the dominant one and $J_Z$ for $Z>1$ do not exceed $\sim 10^{-3} \text{ meV}$ therefore they may be considered as insignificant.

The last type of two-center interactions are correlated hoppings $V$ whose values are presented in Fig.~\ref{fig:figure11}. Intriguingly, $|V_1|$ is of the order of magnitude of $K_1$ and does not behave monotonically in the considered range of $V_0$. Namely, it attains minimum $V_1\approx -0.05\text{ meV}$ for $V_0\approx 2.75 \text{ meV}$. The lack of monotonicity is also present for $V_4$ and $V_5$, however, their values are orders of magnitude smaller than $V_1$.

\begin{figure}
 \includegraphics[width=0.5\textwidth]{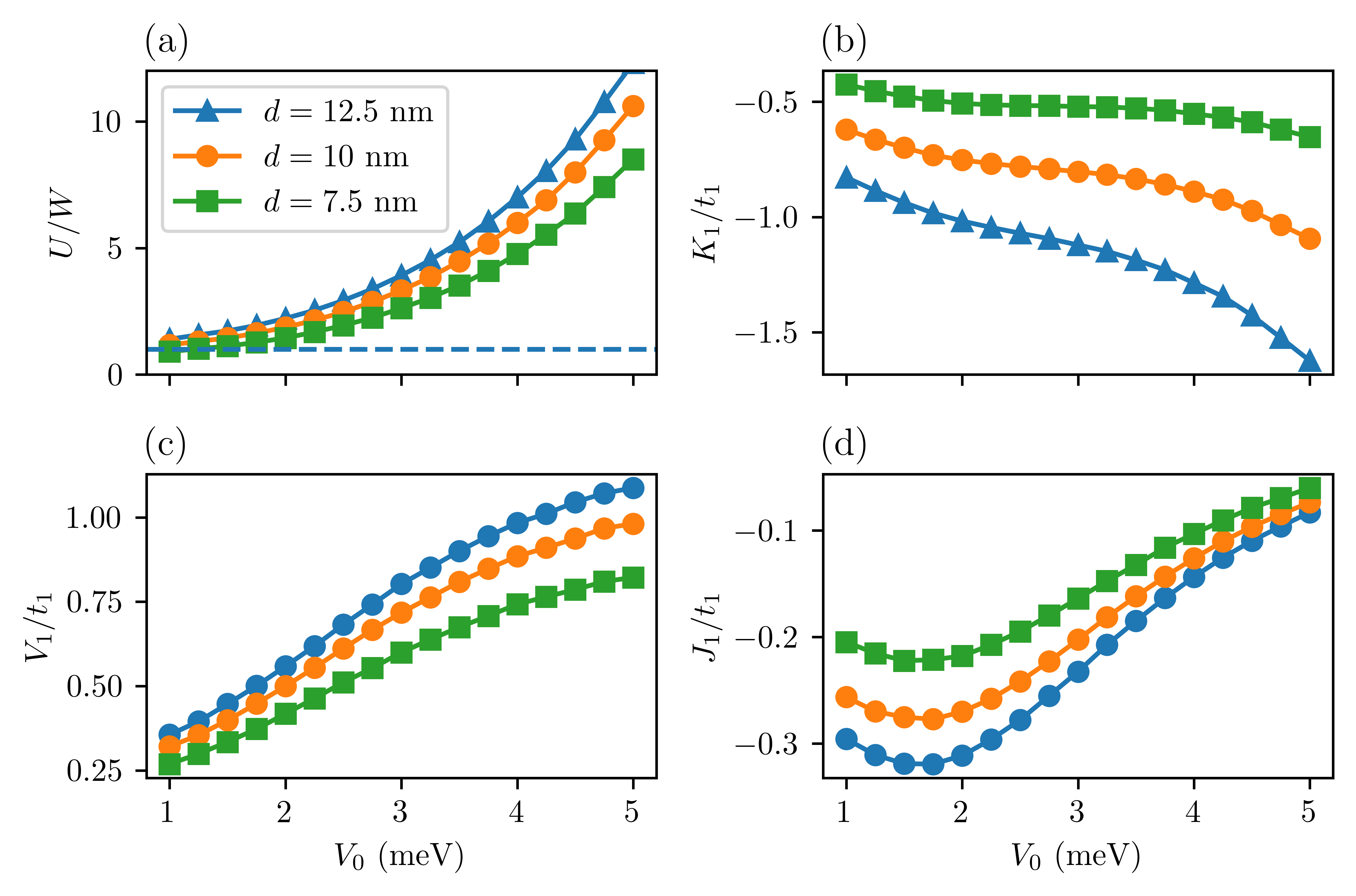}
 \caption{Relations between electron-electron interaction magnitude and kinetic ingredient for the three different regimes of screening tuned by value of $d$. The ratio between $U$ and bare dispersion bandwidth $W\equiv \epsilon(\mathbf{k})_{max}-\epsilon(\mathbf{k})_{min}$ (a); charge density-density $K_1$ related to $t_1$ (b); correlated hopping $V_1$  and exchange interaction $J_1$ divided by $t_1$ (c) and (d) respectively. The dashed line in (a) relates to $U/W=1$, i.e., conventional limit of strong correlations regime.}
 \label{fig:figure12}
\end{figure}

Whereas the parameter $d$ which tunes the screening effect, does not influence the kinetic energy landscape, it changes the magnitude of interactions. Notably, the relation between electron-electron amplitudes and kinetic energy scale governs the role of electronic correlations in the system. In Fig.~\ref{fig:figure12}(a) we depict the conventional measure of correlation strength, which is the ratio between on-site $U$ Hubbard term and a bare dispersion relation width $W$. One finds that for the considered values of  $V_0$ this ratio systematically grows with the depth of the confinement potential and from this point of view a wide range of strong correlation regimes is accessible in the system. Furthermore,  the impact of screening is relatively small when compared to the ratio between density-density interactions $K_1$  and hopping $t_1$ (Fig.~\ref{fig:figure12}(b)). For $1.5\lessapprox V_0\lessapprox 4.5  \text{ meV}$ a plateau appears, which becomes less prominent with decreasing value of $d$. Nevertheless, each depicted value of $d$ provides the magnitude of ratio $|K_1/t_1|\gtrapprox0.5$. This observation suggests that these types of interactions should be taken into account in the many-body interacting Hamiltonian. Although the absolute values referring to the analogous ratios defined for the correlated hopping and exchange amplitudes (see Fig.~\ref{fig:figure12}(c,d)) are substantially smaller, specifically for lower values of $V_0$, they still can be regarded as meaningful at least for $Z=1$. Moreover, they depend weaker on $d$  as compared to $K_1/t_1$.

As the decay of $K,V$ and $J$ with increasing $r_Z$ is of different rate, and, it is a priori difficult to state which terms in Hamiltonian can be safely neglected (especially for such a subtle phenomenon as superconducting pairing), we decided to include all the two-center terms up to $Z=5$. This choice rather overestimates the role of more distanced (i.e., for $Z>1$ ) interaction amplitudes of $V$ and $J$, however, it allows reducing the doubts concerning the role of each type of interaction in the final picture. Eventually, we analyse the Hamiltonian of the form
\begin{equation}
    \mathcal{\hat{H}}_{QDL}=\mathcal{\hat{T}}+\mathcal{\hat{U}}+\mathcal{\hat{K}}+\mathcal{\hat{V}}+\mathcal{\hat{J}},
    \label{eq:interham}
\end{equation}
where $\mathcal{\hat{T}}$ are one body-terms, and, $\mathcal{\hat{U}},\mathcal{\hat{K}},\mathcal{\hat{V}}$ and $\mathcal{\hat{J}}$ refer to the two-body interaction terms corresponding to $U,K,V$ and $J$ respectively for the neighbouring sites corresponding to $Z\leq5$.

Here, we intend to focus on an experimental setup which potentially may reproduce electronic properties similar to those observed in typical strongly correlated materials, i.e., the emergence of antiferromagnetic (AF) order, charge gap development for the half-filled band, and finally the creation of superconducting pairing for the doped cases. Whereas the conventional prerequisite for a strong correlation, i.e., $U/W$ is $\gtrapprox 1$, is already realized in the system, the relations between other interactions and hopping terms may still play an important role. 

\begin{figure}
 \includegraphics[width=0.5\textwidth]{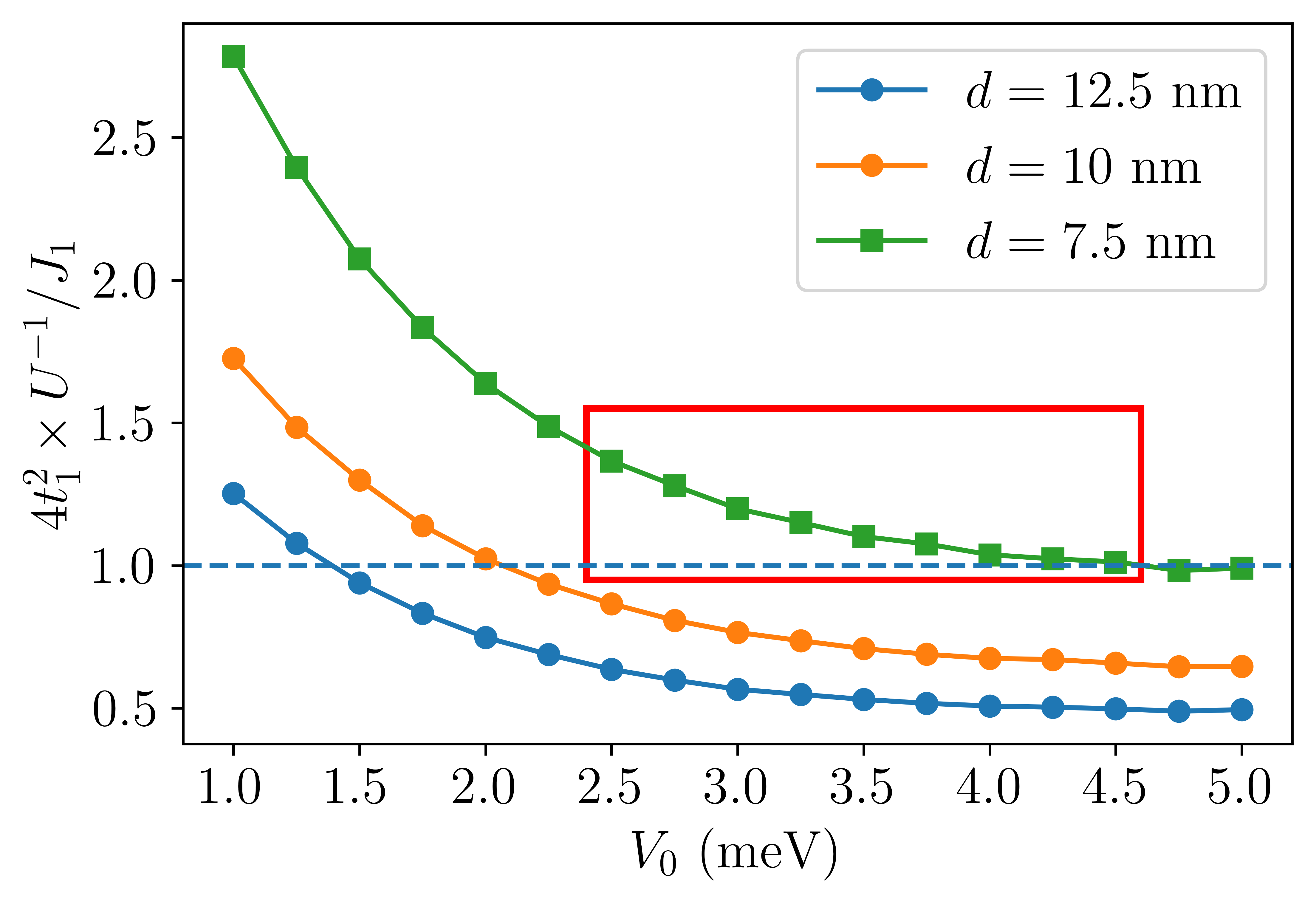}
 \caption{The relation  $(4t_1^{2}/U)/(J_1)$ as a function of $V_0$. The dashed horizontal line separates estimated applicability of Hubbard/$t-j$ (upper) and ferromagnetic Heisenberg (lower) models close to the half-filling. The red rectangle surrounds cases considered in our VMC calculations.}
 \label{fig:figure13}
\end{figure}

Notably, the ferromagnetic interaction $J_1$ may overcome the AF tendency. The latter is an inherent feature of the repulsive Hubbard model at half filling as it can be transformed to the $t-J$ model when double occupancies are projected out~\cite{Spalek_2007}. Therefore, before selecting the set of microscopic parameters to be applied for the variational solution of our Hamiltonian, we carried out the following reasoning. 
The inclination for AF ordering close to half-filling can be estimated by comparing the coupling $J_{t-J}=4t^2/U$ derived for $t-J$ model and the value of the ferromagnetic exchange integral, $J_1$. As shown in Fig. \ref{fig:figure13}, such an estimation suggests that when \emph{screening} is enhanced, i.e., $d$ decreases, the antiferromagnetic order can be expected to dominate. 
Note also, that  $d<L/2$ must hold to satisfy the consistency of model. For $d=7.5 \text{ nm}$ the thickness of layer below the pattern of electrodes which is not penetrated by electrons from 2DEG is still $2.5 \text{ nm}$.

Moreover, the pair-hopping term associated with $J_1$  can lead to Cooper pair modulation of the onsite s-wave paired state \cite{Granath2018}. However, in the regime of relatively large onsite repulsion $U$ the intersite paired state is much more likely to be expected. The mentioned mechanism of inducing the Cooper pair modulation is not going to be operative in such case of intersite pairing. Thus, we turn to a detailed investigation of the homogeneous intersite paired state. 

Finally, we consider the general Hamiltonian given in Eq.(\ref{eq:interham}) for $d = 7.5\text{ nm}$ and $V_0\in[2.0,4.0\text{ meV}]$ which we suspect have properties similar to those of the descendants of the repulsive Hubbard model in the strong correlation regime.

\subsection{VMC solution for the interacting system}
As mentioned in Sec. II the variational ansatz employed in this work can be regarded as highly generic, i.e., capable to describe a variety of complex phases. However, as it is a challenging task from the optimization perspective, even for the less complex model Hamiltonians, we exclude the possibility of long range charge ordering in our analysis assuming $1\times1$ variational parameters sub-lattice\cite{Misawa}. Nevertheless, the adopted form is flexible enough to cover singlet $s-$ and $d-$wave pairing as well as antiferromagnetic spin ordering. Namely, we do not impose any rotational symmetry for $F_{ij}^{\uparrow\downarrow}$ and \emph{Jastrow} type variational parameters. Also, as we consider a single-band model, the final set of variational parameters is augmented with the single \emph{Gutzwiller} projector parameter $g$, as well as, with ten parameters associated with  local electron occupancy configurations supplying single \emph{doublon-holon} correlator.

For a lattice consisting of $N_x\times N_y$  sites the number of  variational parameters scales as the number of considered QDs, since the pattern of connections between sites is repeated for each site, eventually forming a translationally invariant scheme.

\subsubsection{Spin-spin correlation functions and charge gap}
As stated, we are mainly interested if the considered system exhibits properties characteristic for strongly correlated systems such as cuprates. Namely, we expect that for the half-filled band, AF ordering should appear. The analysis of the magnetic properties has been carried out by calculating the correlation functions defined for the  spin $z$-component as
\begin{align}
    S^z(\mathbf{r})=\frac{1}{N_x\times N_y}\sum_{i}\big\langle(\hat{n}_{i\uparrow}-\hat{n}_{i\downarrow})(\hat{n}_{f(\mathbf{R_i}+\mathbf{r})\uparrow}-\hat{n}_{if(\mathbf{R_i}+\mathbf{r})\downarrow})\big\rangle,
\end{align}
where function $f(\mathbf{r})$ maps vector $\mathbf{r}$ onto the proper lattice index. Also,  we consider correlation functions defined in the momentum space which are Fourier transforms of their real-space counter parts  $S^z(\mathbf{r})$, i.e.,
\begin{align}
    S^z(\mathbf{k})=\sum_{j}\text{e}^{i\mathbf{k}\cdot\mathbf{R}_j}S^z(\mathbf{R}_j).
\end{align}

First, we intend to verify if our conjectures regarding the AF ordering at half-filling for the selected range of $V_0$ and $d=7.5\text{ nm}$ are valid. Therefore, we performed preliminary calculations for the system of size $8 \times 8$ and for doping $\delta=0$, where 
\begin{align}
\delta=1-\frac{N_{el}}{N_x \times N_y}.
\end{align}
To reduce the finite size effects, we imposed periodic boundary conditions on the lattice. 
\begin{figure}
 \includegraphics[width=0.5\textwidth]{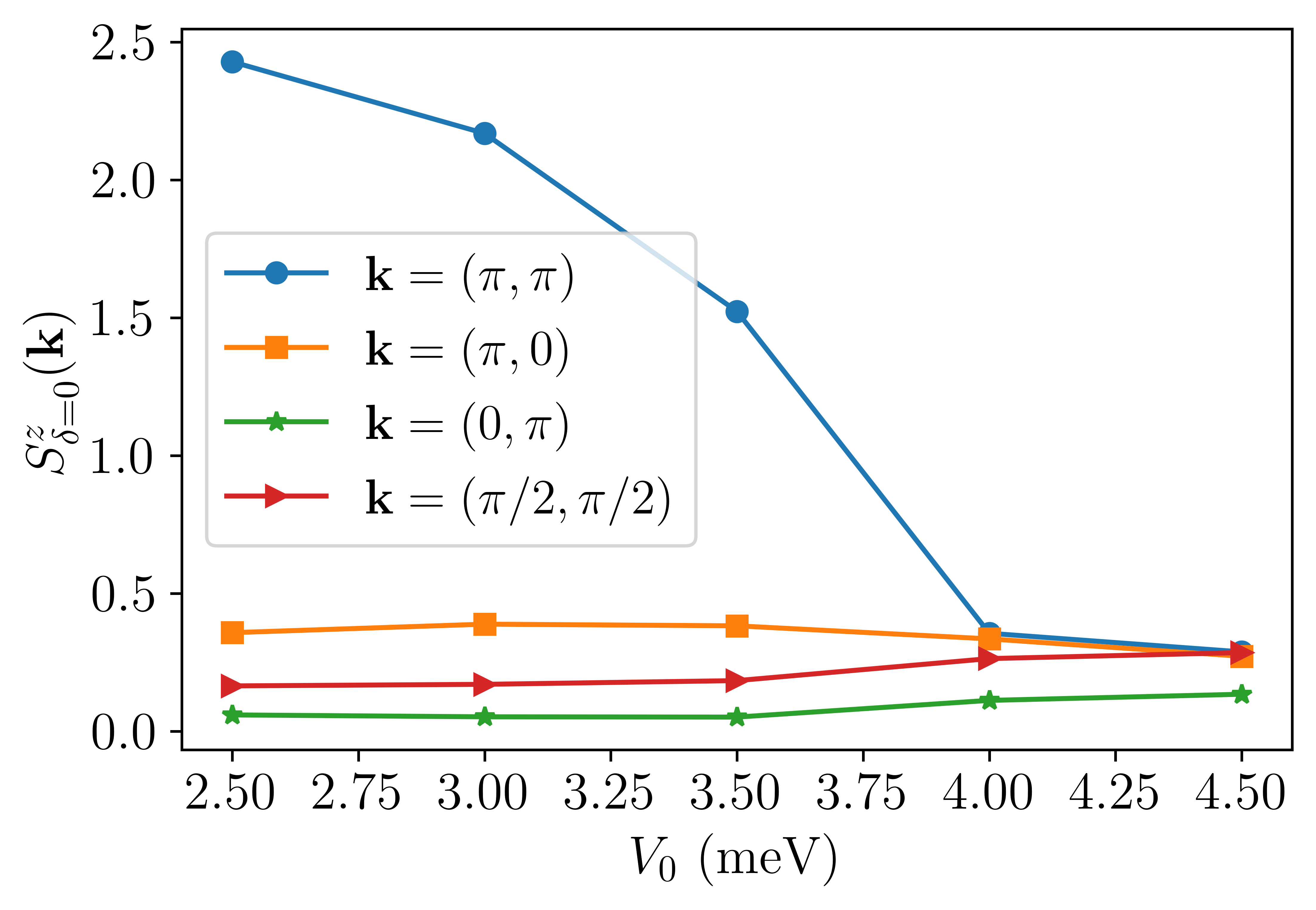}
 \caption{Momentum resolved spin-spin correlation functions for the representative values of $\mathbf{k}$ at $\delta=0$ as a function of $V_0$ for the system consisting of $8\times8$ QDs. }
 \label{fig:figure14}
\end{figure}

The resulting evolution of $S^z(\mathbf{k})$ with $V_0$ for different magnetic modulation vectors is presented in Fig.~\ref{fig:figure14}. It is clearly visible that the antiferromagnetic type of spin ordering dominates for $2.5\lessapprox V_0\lessapprox 3.5 \text{ meV}$. Already here we can see that the ground state magnetic properties of the system can be tuned by changing the gate potential $V_0$.

\begin{figure}
 \includegraphics[width=0.5\textwidth]{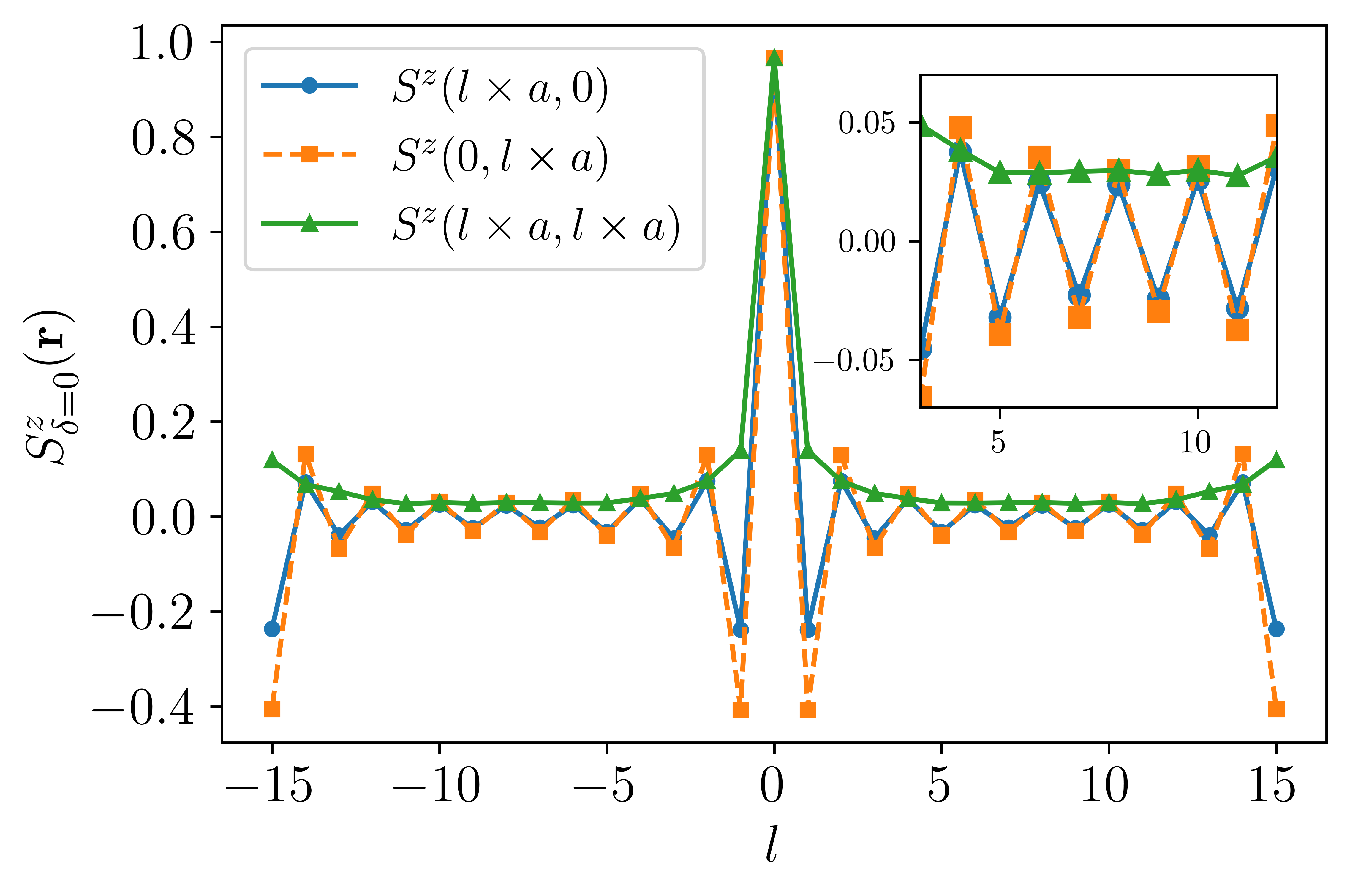}
 \caption{Real-space spin-spin correlation functions at $\delta=0$, for $V_0=3\text{ meV}$ and lattice consisting of $16\times16$ QDs. The discrepancy between $\mathbf{\hat{x}}$ and $\mathbf{\hat{y}}$ directions  is observed, however this difference decays with increasing $|l|$. The statistical error is smaller than symbols size.}
 \label{fig:figure15}
\end{figure}

In Figure~\ref{fig:figure15} we show how the spin-spin correlation function changes in real space along the $(1,0)$, $(0,1)$, and $(1,1)$ directions of the QD square lattice for a larger system ($16\times16$) and for the gate potential $V_0=3$ meV. In agreement with the results presented in Fig.~\ref{fig:figure14} we observe the behavior characteristic for the AF ordering. It should be noted that small differences between the $(1,0)$ and $(0,1)$ directions are visible in Fig.~\ref{fig:figure15} in spite of the fact that the system itself is $C_4$ symmetric. Such a spontaneous symmetry breaking is possible since the exploited variational ansatz is not constrained to be symmetric with respect to the spatial directions. Despite these circumstances, the AF ordering is strongly manifested. In addition, the $(1,0)$/$(0,1)$ asymmetry is decreasing with increasing distance as well as it is also weaker as the system is becoming larger. Therefore, the observed $C_4$ symmetry breaking can vanish in the thermodynamic limit for the considered variational ansatz.

\begin{figure}
 \includegraphics[width=0.5\textwidth]{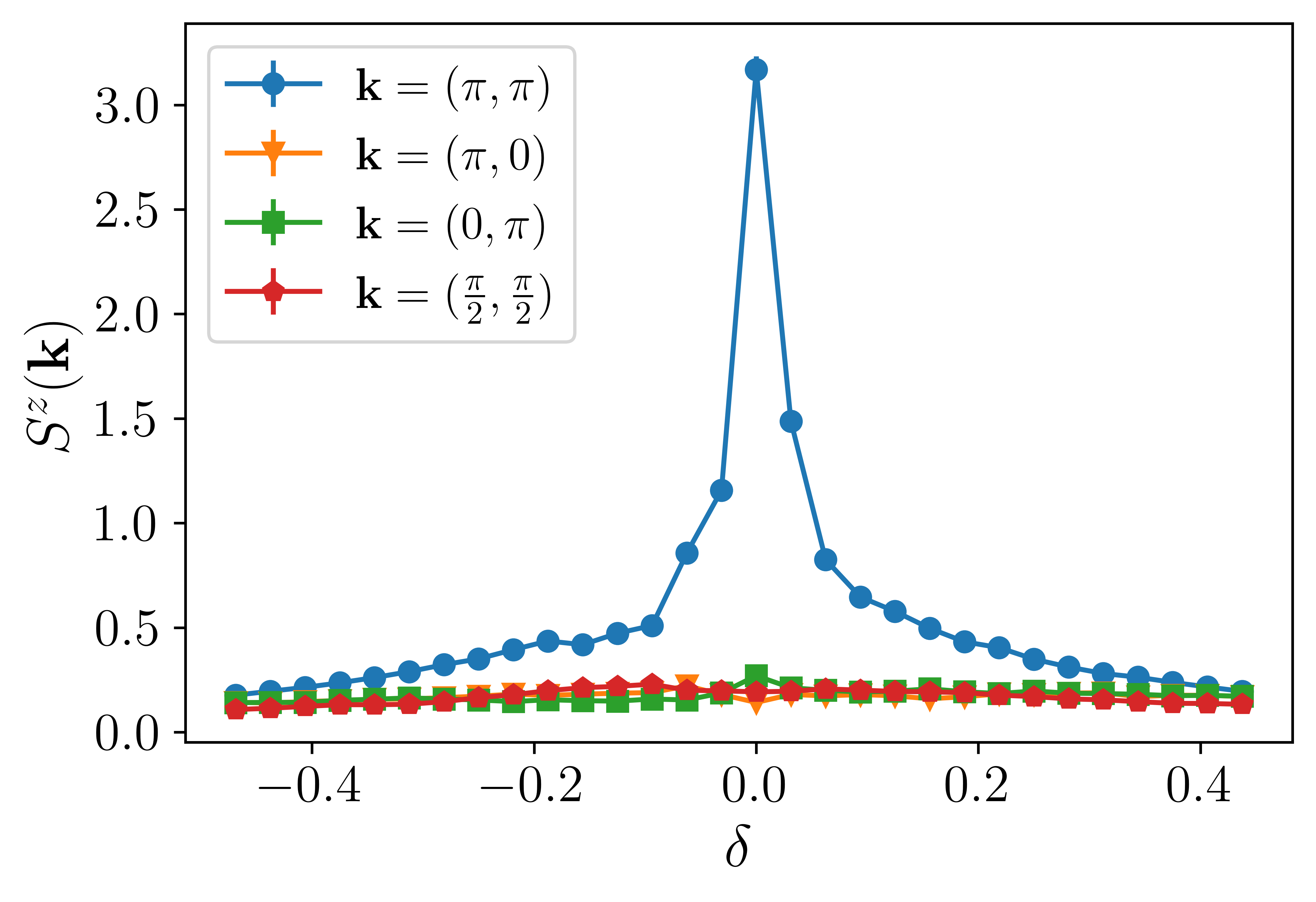}
 \caption{Momentum space spin-spin correlation functions for $V_0=3\text{ meV}$ and lattice consisting of $16\times16$ QDs as a function of doping.}
 \label{fig:figure16}
\end{figure}

For the sake of completeness in Fig.~\ref{fig:figure16} we present $S^z(\mathbf{k})$ as a function of both electron and hole doping $\delta\in[-0.5,0.5]$ for $\mathbf{k}\in\{(\pi,\pi),(0,\pi),(\pi,0),(\pi/2,\pi/2)\}$. As expected, AF ordering dominates. For one electron per single QD, a well-pronounced peak emerges and the values corresponding to the remaining directions are marginal. 

\begin{figure}
 \includegraphics[width=0.5\textwidth]{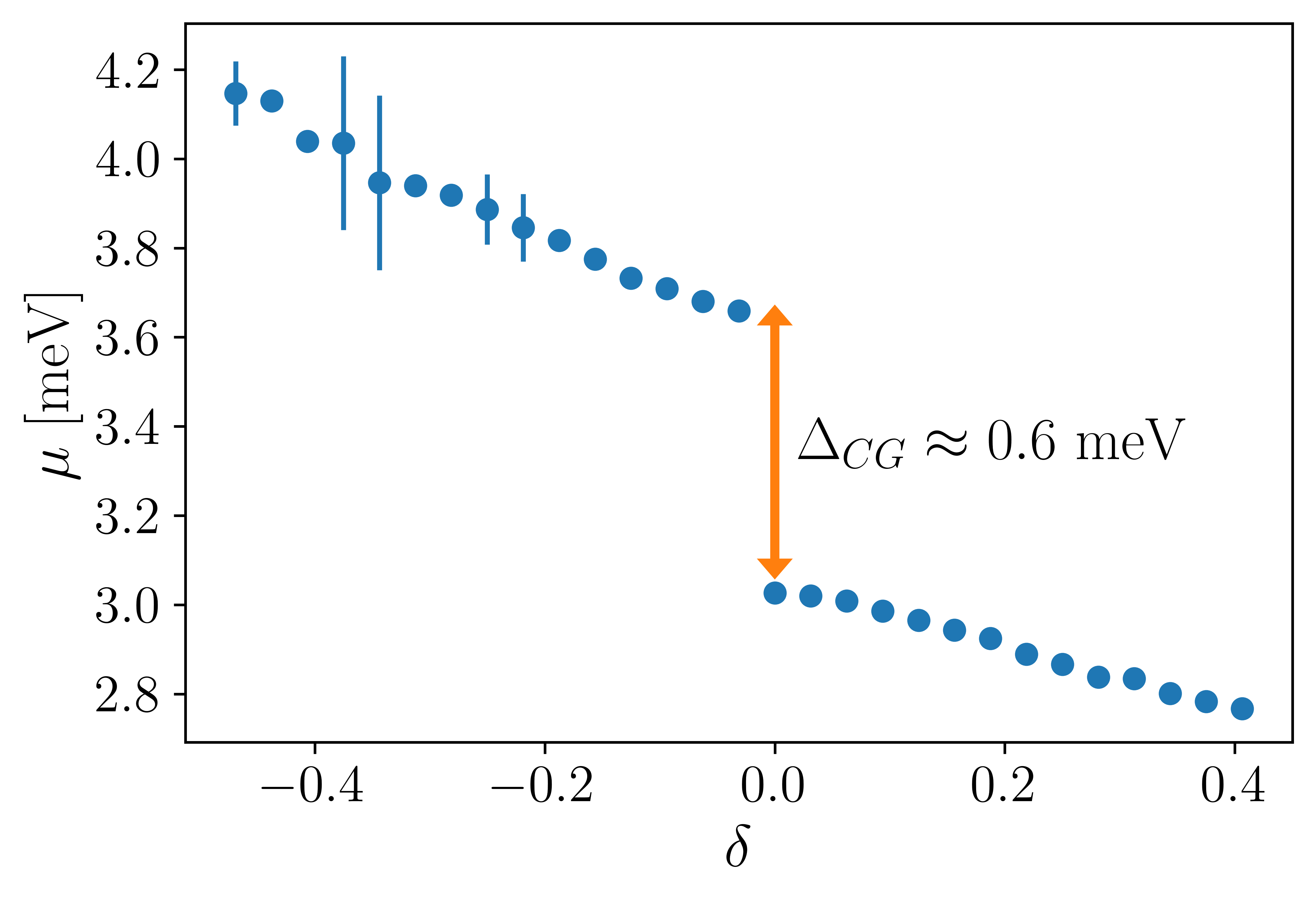}
 \caption{Estimated chemical potential $\mu$ as a function $\delta$. The arrow represent resulting  charge-gap estimation. In most of cases, errors are smaller than the symbol size.}
 \label{fig:figure17}
\end{figure}

The appearance of the AF ordering close to the zero-doped case is typical for the family of copper-based high-temperature superconductors\cite{Ogata_2008}. For the latter systems, the $(\pi,\pi)$ magnetic modulation is accompanied by the electron-electron interaction induced insulating state. Therefore, here we also analyze the possibility of charge gap creation which is a signature of the Mott insulator. For this reason, we calculate the approximate value of chemical potential
\begin{align}
\mu(\delta)\approx\frac{E(N_{el})-E(N_{el}-\Delta N_{el})}{\Delta N_{el}},
\end{align}
which allows us to determine the value of the charge gap in the following manner
\begin{align}
\Delta_{CG}=\lim_{\delta\rightarrow 0^-}\mu(\delta)-\lim_{\delta\rightarrow 0^+}\mu(\delta).
\end{align}
In Fig. \ref{fig:figure17} we present $\mu(\delta)$ for $16\times16$ QD lattice within resolution $\Delta N_{el}=8$. An abrupt change of the chemical potential for $\delta=0$, resulting in $\Delta_{CG}\approx0.6\text{ meV}$ indicates an insulating character of the system. 
Since in our case the Hubbard interaction is $U\approx1.2\text{ meV}$ and the bandwidth is $W\approx0.45\text{ meV}$, the simplest estimate of the charge-gap for the case of the Hubbard model can be given roughly as $\Delta_{CG}=U-W\approx0.75\text{ meV}$---a value which is close to the one obtained here. We find that the principal features of the considered system are determined by the intra-site repulsion together with the form of band structure, which is the case known from the Hubbard model.

It should be noted that in general the intersite Coulomb repulsion terms may induce the insulating state at the $\delta=\pm0.5$ dopings and/or the charge ordered states. However, as shown in Fig. 8, the intersite Coulomb repulsion integrals are relatively small in the considered parameter range. Therefore, the mentioned effects are not expected here.

Another prominent phenomenon related to the physics of strongly correlated systems is the formation of the superconducting state upon electron or hole doping. The appearance of the latter in the considered system is analyzed in the following subsection.

Since even for the pure Hubbard model, the full phase diagram is not indisputably settled \cite{arovas,Qin2}, one may expect that for the more complex Hamiltonian considered here, a variety of charge and spin ordered states should be put under examination. This is, however, beyond the scope of the present study, which focuses on \emph{archetypal} properties of strongly correlated systems in view of Hubbard-like models.

\subsubsection{Pairing correlation functions}
We analyze the possibility of singlet pairing by considering the pair correlation functions defined in the following manner
\begin{equation}
    P_{q}(\mathbf{R}_{ij})=\sum_{\mathbf{a},\mathbf{a}'}g_q(\mathbf{a})g_q(\mathbf{a}')\big\langle\hat{\Delta}^{\dagger}_{i,f(\mathbf{R_i}+\mathbf{a})}\hat{\Delta}_{j,f(\mathbf{R_j}+\mathbf{a}')}\big\rangle,
\end{equation}
where $q$ stands for pairing channel ($d$ or $s$), $\mathbf{a}$ ($\mathbf{a}'$) is the $(\pm a,0),(0,\pm a)$ vector, and $g_q(\mathbf{a})$ is the symmetry-related factor which takes the form $g_s(\mathbf{a})=g_d(\pm a,0)=-g_d(0,\pm a)=1$. The singlet-pairing operators $\hat{\Delta}^{\dagger}_{ij}$ are defined as
\begin{equation}
    \hat{\Delta}^{\dagger}_{ij}=\frac{1}{\sqrt{2}}\big(\hat{c}_{i\uparrow}^{\dagger}\hat{c}_{j\downarrow}^{\dagger}-\hat{c}_{i\downarrow}^{\dagger}\hat{c}_{j\uparrow}^{\dagger}\big).
\end{equation}

As the tendency for pairing should be analyzed for $\lim_{|\mathbf{R_{ij}}| \to \infty}P_q(\mathbf{R}_{ij})$, what cannot be realized in practice for a finite lattice, non-zero values of $P_q(\mathbf{R}_{ij})$ can be falsely interpreted as signatures of the superconducting state\cite{White}. To overcome this difficulty, we compute the vertex pairing functions\cite{White,Kuroki,Fang,HuangLin} given as
\begin{equation}
 \begin{split}
 \overline{P}_{q}(\mathbf{R}_{ij})=P_{q}(\mathbf{R}_{ij})-\sum_{\mathbf{a},\mathbf{a}'}g_q(\mathbf{a})g_q(\mathbf{a}') \\ \times \Big[\langle \hat{c}_{i\uparrow}^{\dagger}\hat{c}_{j\uparrow}\rangle \langle \hat{c}_{f(\mathbf{R_i}+\mathbf{a})\downarrow}^{\dagger}\hat{c}_{f(\mathbf{R_j}+\mathbf{a}')\downarrow}\rangle \\+\langle \hat{c}_{i\uparrow}^{\dagger}\hat{c}_{\mathbf{R_j}+\mathbf{a}'\uparrow}\rangle \langle \hat{c}_{f(\mathbf{R_i}+\mathbf{a})\downarrow}^{\dagger}\hat{c}_{j\downarrow}\rangle\\
 +\langle \hat{c}_{i\downarrow}^{\dagger}\hat{c}_{j\downarrow}\rangle \langle \hat{c}_{f(\mathbf{R_i}+\mathbf{a})\uparrow}^{\dagger}\hat{c}_{f(\mathbf{R_j}+\mathbf{a}')\uparrow}\rangle \\+\langle \hat{c}_{i\downarrow}^{\dagger}\hat{c}_{\mathbf{R_j}+\mathbf{a}'\downarrow}\rangle \langle \hat{c}_{f(\mathbf{R_i}+\mathbf{a})\uparrow}^{\dagger}\hat{c}_{j\uparrow}\rangle \Big].
 \end{split}
\end{equation}

\begin{figure}
 \includegraphics[width=0.5\textwidth]{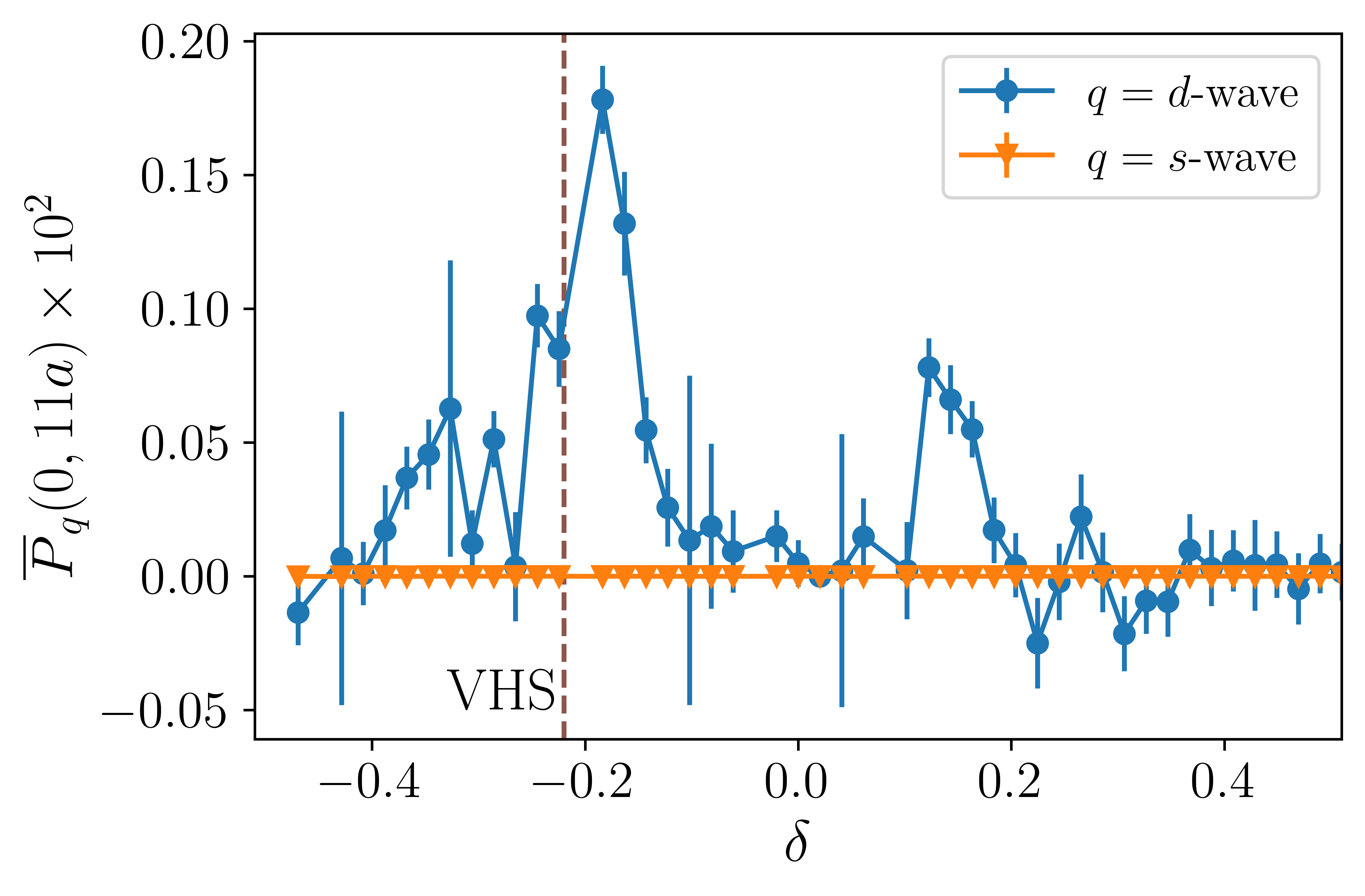}
 \caption{Pairing $\overline{P}_{q}$ at maximal possible distance $\mathbf{R}_{max}$ for $8\times24$ QD lattice as a function of doping. Dashed horizontal line refers to van Hove singularity obtained for the non-interacting system.}
 \label{fig:figure18}
\end{figure}

The subtraction of the products of the hopping averages allows finding a reliable estimation of the tendency for the pairing in the finite size system\cite{White}.
We obtain the vertex functions for $\mathbf{R}_{ij}$ oriented along $\mathbf{\hat{x}}$ for $V_0=3 \text{ meV}$, $d=7.5 \text{ nm}$ and for the lattice size $L_x\times L_y = 8\times24$. In Fig.~\ref{fig:figure18} we present $\overline{P}_{q}$ for both $d-$ and $s-$wave pairing symmetries as a function of doping $\delta$, calculated for $R_{max}=(0,11a)$. As the pairing amplitude in general is relatively small and may be affected by statistical noise, the averaging procedure is performed within $\sim10^8$ Monte-Carlo steps and repeated ten times for each considered doping to estimate the statistical error.

As shown in Fig.~\ref{fig:figure18} a dome-like behavior for the $d$-wave amplitude appears in the electron-doped region of the phase diagram. Namely, non-zero vertex pairing functions are obtained for $-0.37\lessapprox\delta\lessapprox-0.12$, as well as, for the hole doped case $0.1\lessapprox\delta\lessapprox0.18$. The latter exposes a noticeable smaller amplitude than the former and poses a less smooth shape. Nevertheless, both doping regimes show a clear tendency towards the superconducting phase formation within the $d$-wave pairing channel.

\begin{figure}
 \includegraphics[width=0.5\textwidth]{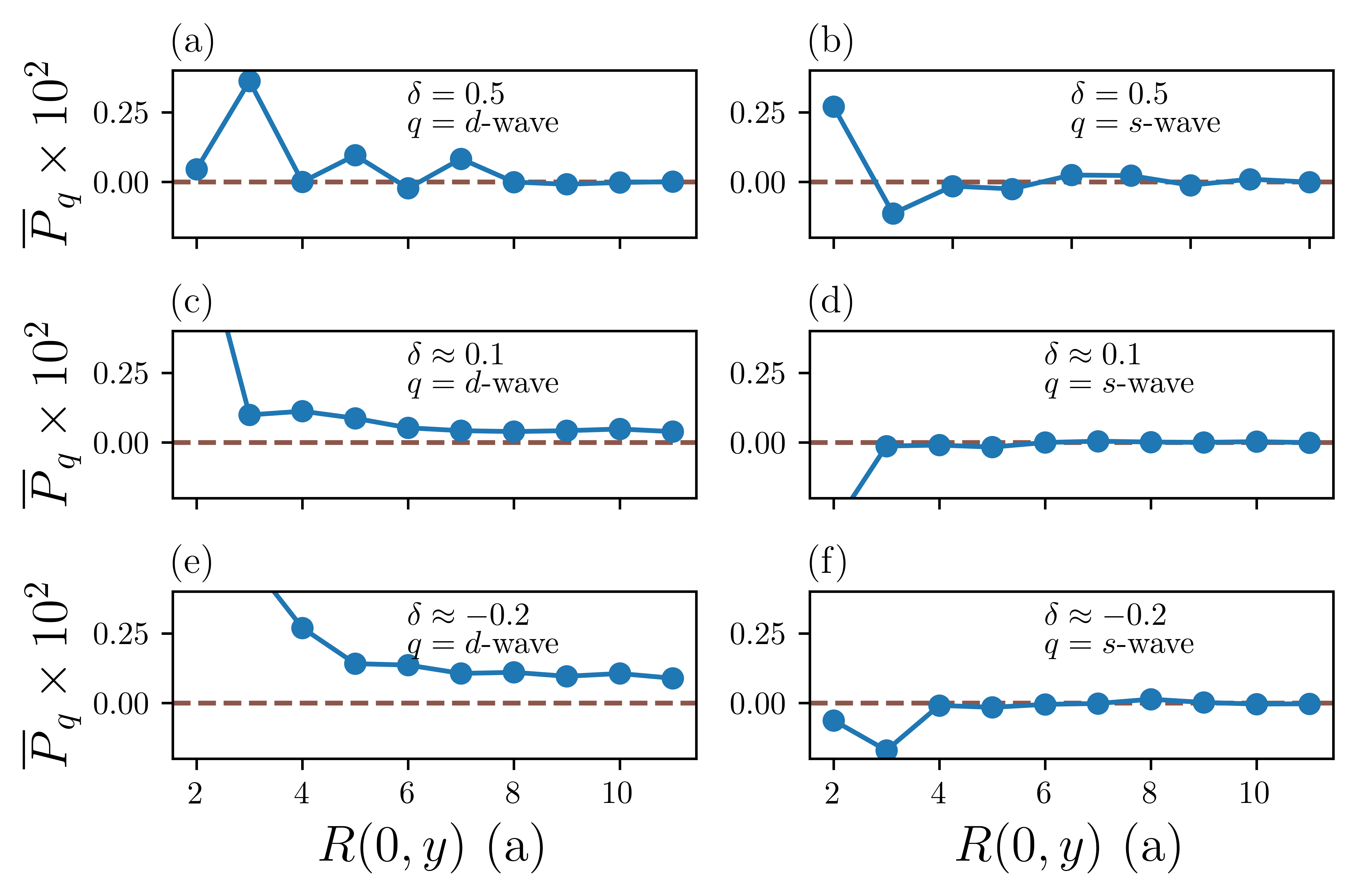}
 \caption{Pairing $\overline{P}_{q}$ as function of distance $R(0,y)$  for $8\times24$ QDL  for the selected values of doping $\delta$.}
 \label{fig:figure19}
\end{figure}

In Fig.~\ref{fig:figure19} the spatial dependency of $\overline{P}_{q}$ is shown for three representative cases. We find that the chosen system dimensions allow capturing the asymptotic tendency. Namely, for $R(0,y)\gtrapprox8$ pairings related to $d$-wave channel converge within the estimated statistical error. The $s$-wave pairing vertex function decays to the value close to zero even more rapidly, i.e., for $R(0,y)\gtrapprox4$ only the residual values are observed. Furthermore, we disregard the analysis of inconclusive cases, i.e., those for which related statistical errors or/and values are greater of the order of magnitude than those exhibited by majority.

The two peaks related to the $d$-wave pairing regions separated by the AF phase resemble the phase diagram identified for the cuprate superconductors. However, in our case, the paired state is more pronounced at the electron-doped site instead of the hole-doped situation. This can be understood in view of the so-called van Hove singularity scenario\cite{Bok} according to which the peak in the density of states for the layered structures has a significant influence on the enhancement of the paired state regardless of the particular pairing mechanism. Although the BCS-type mechanism is not often discussed in view of the cuprate physics and is surely absent in the QD lattice analyzed here, the peak in the density of states may also influence the correlation-driven pairing as it is suggested in Ref. \onlinecite{Markiewicz}. Since for the cuprates the van Hove singularity is reached at the hole-doped site of the phase diagram, the superconducting dome is more pronounced for $\delta>0$. On the other hand, for the case analysed here, the bare band-structure calculations reveal the singularity at $\delta\approx-0.22$, i.e., inside electron-doped regime. As shown in Fig.~\ref{fig:figure18}, this value coincides with the doping range for which the obtained amplitudes of $d$-wave pairing are characterized by the largest magnitude. Additionally, close to $\delta=0$ the emergence of AF state appears (see Fig.~\ref{fig:figure16}) together with the Mott insulating state, what leads to the suppression of the paired state in the close proximity of half-filling and leaving an asymmetric two-dome structure of the superconducting pairing amplitudes. 

Finally, as shown the $d$-$wave$ symmetry dominates over the $extended$ $s$-$wave$ pairing in the considered doping range. This fact comes as a result of the relative position of the Fermi surface and the nodal lines corresponding to the particular symmetry factor. The symmetry that results in less suppression of the SC gap at the Fermi surface is chosen by the system as it makes the paired state more stable. The obtained result is in agreement with the previous calculations made for the simple Hubbard model where the d-wave symmetry is also stable in the range around the half-filled situation\cite{Kaczmarczyk2013}.


\section{Summary}
We have developed a model describing the QD lattice and numerically analysed the electronic characteristics in the strongly correlated regime within the low carrier density regime.

We have shown that the system can be described by a single band of interacting electrons. We have found that by properly tuning the gate potential, one can reach the situation in which the width of the band is relatively small when compared to the electron-electron interaction amplitudes and thus making the properties of the system dominated by electronic correlations. Furthermore, by estimating the antiferromagnetic coupling within the approach based on the $t$-$J$ model we have found that the inherent correlation-driven effective antiferromagnetic ordering can overcome the trend of ferromagnetic spin alignment.

The Variational Monte-Carlo simulations for the obtained model have shown that the QD lattice exhibits three main features of strongly correlated high-temperature superconductors: (\emph{i}) tendency towards AF ordering when the charge density is close to one electron per QD: (\emph{ii}) development of charge gap which can be related to the Mott insulating state, and, (\emph{iii}) emergence of the $d$-wave pairing which exhibits a two-dome structure. We have found that the pairing is stronger in the electron-doped regime, in contrast to the situation observed in cuprates. This can be related to the character of the bare band structure, which exhibits a peak in the density of states for the case of electron-doped case. 

Therefore, one may conclude that the main features of the system under consideration qualitatively resemble those known from cuprates, rendering such a QD lattice as an experimentally accessible, electrically tunable, artificial material that would allow studying the physics of correlated systems. 

\section{Acknowledgement}
This work was supported by National Science Centre (NCN) agreement number UMO-2020/38/E/ST3/00418.

\appendix*
\section{Microscopic parameters for the interacting Hamiltonian}
\label{Appendix}
In this Appendix we provide values of microscopic parameters up to $Z=5$ for $V_0=3.0\text { meV}$ and $d=7.5\text{ nm}$, i.e., for the system for which the analysis of magnetic as well as pairing properties has been carried out for the interacting Hamiltonian. An estimated error is $\propto10^{-4}$. 

In addition, the code and the data behind the presented Figures can be downloaded from an open repository together with the values of the adopted microscopic parameters\cite{andrzej_biborski_2021_5529809}.
\begin{table}[h!]
\centering
\begin{tabular}{ |p{1.3cm}|p{1.3cm}|p{1.3cm}|p{1.3cm}|p{1.3cm}|p{1.3cm}|  }
 \hline
 \multicolumn{6}{|c|}{Single particle parameters (meV)} \\
 \hline\hline
 $\epsilon$    & $t_1$&  $t_2$ & $t_3$ & $t_4$ & $t_5$\\
 \hline
 $2.657$    & $-0.061$&  $-0.002$ & $0.007$ & $0.001$ & $0.000$\\
 \hline
\end{tabular}
\caption{Values of single particle parameters, note that $\epsilon\equiv t_0$.}
\end{table}

\begin{table}[h!]
\centering
\begin{tabular}{ |p{1.3cm}|p{1.3cm}|p{1.3cm}|p{1.3cm}|p{1.3cm}|p{1.3cm}|  }
 \hline
 \multicolumn{6}{|c|}{Density-density Coulomb interactions (meV)} \\
 \hline\hline
 $U$    & $K_1$&  $K_2$ & $K_3$ & $K_4$ & $K_5$\\
 \hline
 $1.262$    & $0.032$&  $ 0.006$ & $0.002$ & $ 0.001$ & $0.001$\\
 \hline
\end{tabular}
\caption{Values of Coulomb density-density interaction amplitudes, here $U\equiv K_0$.}
\end{table}

\begin{table}[h!]
\centering
\begin{tabular}{ |p{1.3cm}|p{1.3cm}|p{1.3cm}|p{1.3cm}|p{1.3cm}|  }
 \hline
 \multicolumn{5}{|c|}{Exchange interactions (meV)} \\
 \hline\hline
 $J_1$&  $J_2$ & $J_3$ & $J_4$ & $J_5$\\
 \hline
 $0.010$    & $0.000$&  $ 0.000$ & $0.000$ & $ 0.000$\\
 \hline
\end{tabular}
\caption{Values of exchange interaction amplitudes.}
\end{table}
\begin{table}[h!]
\centering
\begin{tabular}{ |p{1.3cm}|p{1.3cm}|p{1.3cm}|p{1.3cm}|p{1.3cm}|  }
 \hline
 \multicolumn{5}{|c|}{Correlated hoppings interactions (meV)} \\
 \hline\hline
 $V_1$&  $V_2$ & $V_3$ & $V_4$ & $V_5$\\
 \hline
 $-0.037$    & $0.004$&  $ 0.001$ & $0.000$ & $ 0.000$\\
 \hline
\end{tabular}
\caption{Values of correlated hopping interaction amplitudes.}
\end{table}

\bibliography{biborski_et_al}

\end{document}